\documentclass[authoryear]{aastex}
\usepackage{emulateapj5}
\usepackage{epic} 
\usepackage{float} 
\usepackage{amsmath, amssymb} 
\input epsf

\newcommand{\myfig} {Figure}
\newcommand{\mytab} {Table} 
\newcommand{\wfu} {$U_{300}$}
\newcommand{\wfb} {$B_{450}$}
\newcommand{\wfv} {$V_{606}$}
\newcommand{\wfi} {$I_{814}$}
\newcommand{\vltj} {$J_s$}
\newcommand{\vlth} {$H$}
\newcommand{\vltk} {$K_s$}

\newcommand{\cjk} {$J_s - K_s$}

\newcommand{\cik} {$I_{814} - K_s$}
\newcommand{\civ} {$I_{814} - V_{606}$}
\newcommand{\hdfslong} {Hubble Deep Field South}
\newcommand{\hdfs} {HDF-S}
\newcommand{\fireslong} {Faint InfraRed Extragalactic Survey}
\newcommand{\fires} {FIRES}
\newcommand{\isaaclong}{Infrared Spectrometer And Array Camera}

\newcommand{\fireswww} {\url{http://www.strw.leidenuniv.nl/\~{}fires}}
\newcommand{\jtime} {33.6}
\newcommand{\htime} {32.3}
\newcommand{\ktime} {35.6}

\newcommand{\jlimthree} {26.8} 
\newcommand{\hlimthree} {26.2} 
\newcommand{\klimthree} {26.2}

\newcommand{\catkcounts} {833}
\newcommand{\starcounts} {57}

\newcommand{\wmincounts} {624} 
\newcommand{\nphotz} {567}

\shorttitle{Ultradeep Near-Infrared ISAAC Observations of the \hdfs}
\shortauthors{Labb\'e et al.}
\slugcomment{}

\begin{document}    

\title{Ultradeep Near-Infrared ISAAC Observations of the  \\
  Hubble Deep Field South:\\ Observations, Reduction, Multicolor Catalog, and
  Photometric Redshifts\altaffilmark{1}
}

\author{Ivo Labb\'{e}\altaffilmark{2}, Marijn Franx\altaffilmark{2},
Gregory Rudnick\altaffilmark{3}, Natascha M. F\"{o}rster
Schreiber\altaffilmark{2}, Hans-Walter Rix\altaffilmark{4}, Alan
Moorwood\altaffilmark{5}, Pieter G. van Dokkum\altaffilmark{6}, Paul
van der Werf\altaffilmark{2}, Huub R\"{o}ttgering\altaffilmark{2}, 
Lottie van Starkenburg\altaffilmark{2}, Arjen van de Wel\altaffilmark{2}, 
Konrad Kuijken\altaffilmark{2}, Emanuele Daddi\altaffilmark{5}}


\par
\begin{center}
{\em Accepted for publication in the Astronomical Journal, March 2003, vol. 125, nr. 3}
\end{center}

\altaffiltext{1}{Based on service mode observations collected at 
the European Southern Observatory, Paranal, Chile 
(ESO Programme 164.O-0612). Also based on observations with the 
NASA/ESA {\em Hubble Space Telescope}, obtained at the Space 
Telescope Science Institute which is operated by AURA, Inc., 
under NASA contract NAS5-26555.}

\altaffiltext{2}{Leiden Observatory, P.O. Box 9513, NL-2300 RA,
Leiden, The Netherlands}

\altaffiltext{3}{Max-Plank-Institut f\"ur Astrophysik, P.O. Box 1317,
D-85741, Garching, Germany}

\altaffiltext{4}{Max-Plank-Institut f\"ur Astronomie, D-69117,
Heidelberg, Germany }

\altaffiltext{5}{European Southern Observatory, D-85748, Garching,
Germany }

\altaffiltext{6}{California Institute of Technology, MS 105-24,
Pasadena CA 91125, USA}

\begin{abstract}
We present deep near-infrared (NIR) \vltj, \vlth, and \vltk-band ISAAC
imaging of the WFPC2 field of the \hdfslong\ (\hdfs). The $2.5\arcmin
\times 2.5\arcmin$ high Galactic latitude field was observed with the
VLT under the best seeing conditions with integration times amounting
to \jtime\ hours in \vltj, \htime\ hours in \vlth, and \ktime\ hours
in \vltk. We reach total AB magnitudes for point sources of
\jlimthree, \hlimthree, and \klimthree\ respectively ($3\sigma$),
which make it the deepest ground-based NIR observations to date, and
the deepest \vltk-band data in any field. The effective seeing of the
coadded images is $\approx0\farcs45$ in $J_s$, $\approx0\farcs48$ in
$H$, and $\approx0\farcs46$ in $K_s$. Using published WFPC2 optical
data, we constructed a \vltk-limited multicolor catalog containing
\catkcounts\ sources down to $K_{s,AB}^{tot} \lesssim 26$, of which
\wmincounts\ have seven-band optical-to-NIR photometry. These data
allow us to select normal galaxies from their rest-frame optical
properties to high redshift ($z \lesssim 4$). The observations, data
reduction and properties of the final images are discussed, and we
address the detection and photometry procedures that were used in
making the catalog. In addition, we present deep number counts, color
distributions and photometric redshifts of the \hdfs\ galaxies. We
find that our faint $K_s$-band number counts are flatter than
published counts in other deep fields, which might reflect cosmic
variations or different analysis techniques. Compared to the HDF-N, we
find many galaxies with very red $V-H $ colors at
photometric redshifts $1.95 < z_{phot} < 3.5$. These galaxies are bright
in $K_s$ with infrared colors redder than $J_s-K_s > 2.3$ (in Johnson
magnitudes).  Because they are extremely faint in the
observed optical, they would be missed by ultraviolet-optical
selection techniques, such as the U-dropout method.
\end{abstract}

\par


\maketitle
\section{Introduction}

In the past decade, our ability to routinely identify and
systematically study distant galaxies has dramatically advanced our
knowledge of the high-redshift universe. In particular, the efficient
U-dropout technique \citep{S96a,S96b} has enabled the selection of
distant galaxies from optical imaging surveys using simple photometric
criteria. Now more than 1000  of these Lyman break galaxies
(LBGs) are spectroscopically confirmed at $z \gtrsim 2$, and have been
subject to targeted studies on spatial clustering \citep{Gi01},
internal kinematics \citep{Pe98,Pe01}, dust properties \citep{A00},
and stellar composition \citep{Sh01,Pa01}. Although LBGs are among the
best studied classes of distant galaxies to date, many of their
properties like their prior star formation history, stellar population
ages, and masses are not well known.  \par 

More importantly, it is unclear if the ultraviolet-optical selection
technique alone will give us a fair census of the galaxy population at
$z \sim 3$ as it requires galaxies to have high far-ultraviolet
surface brightnesses due to on-going spatially compact and relatively
unobscured massive star formation. We know that there exist highly
obscured galaxies, detected in sub-mm and radio surveys \citep{Sm00},
and   optically faint hard X-ray sources
\citep{Co01,B01} at high redshift that would not be selected as LBGs,
but their number densities are low compared to LBGs and they might
represent rare populations or transient evolutionary phases. In
addition, the majority of present-day elliptical and spiral galaxies,
when placed at $z \sim 3$, would not satisfy any of the current
selection techniques for high-redshift galaxies. Specifically, they
would not be selected as U-dropout galaxies because they are too faint
in the rest-frame UV. It is much easier to detect such galaxies in the
near-infrared (NIR), where one can access their rest-frame optical
light. 
 Furthermore,
observations in the near-infrared allow the comparison of galaxies of
different epochs at fixed rest-frame wavelengths where long-lived
stars may dominate the integrated light. Compared to the rest-frame
far-UV, the rest-frame optical light is less sensitive to the effects
of dust extinction and on-going star formation, and provides a better
tracer of stellar mass.  By selecting galaxies in the near-infrared
\vltk-band, we expect to obtain a more complete census of the galaxies
that dominate the stellar mass density in the high-redshift universe,
thus tracing the build-up of stellar mass directly.

\par
In this context we initiated the \fireslong\ \citep[\fires;][]{Fr00},
 a large public program carried out at the {\em Very Large Telescope} 
 (VLT) consisting of very deep NIR imaging of two selected fields. We
  observed fields with existing deep optical WFPC2 imaging from the 
  {\em Hubble Space Telescope} (HST): the WPFC2-field of 
  \hdfslong\ (\hdfs), and the field around the $z\approx0.83$ cluster 
  MS1054-03. The addition of NIR data to the optical photometry is 
  required not only to access the rest-frame optical, but also to 
  determine the redshifts of faint galaxies from their broadband 
  photometry alone. While it may be possible to go to even redder 
  wavelengths from the ground, the gain in terms of effective wavelength leverage is 
  less dramatic compared to the threefold increase going from the 
  $I$ to $K$-band. This is because the \vltk-band is currently the
   reddest band where achievable sensitivity and resolution are 
   reasonably comparable to deep space-based optical data. Preliminary 
   results from this program were presented by \citet[][ hereafter R01]{Ru01}.
\par 

Here we present the full NIR data set of the HDF-S, together with a
\vltk-selected multicolor catalog of sources in the \hdfs\ with
seven-band optical-to-infrared photometry (covering $0.3 -
2.2\micron$), unique in its image quality and depth. This paper
focusses on the observations, data reduction and characteristic
properties of the final images. We also describe the source detection
and photometric measurement procedures and lay out the contents of the
catalog, concluding with the NIR number counts, color distributions of
sources, and their photometric redshifts.  The results of the
MS1054-03 field will be presented by \citet{Fo02} and a more detailed
explanation of the photometric redshift technique can be found in
\citet{Ru01,Ru02b}.  
Throughout this paper, all
magnitudes are expressed in the AB photometric system \citep{Ok71}
unless explicitly stated otherwise.

\section{Observations}
\subsection{Field Selection and Observing Strategy}
The high Galactic latitude field of the \hdfs\ is a natural choice for
follow-up in the near-infrared given the existing ultradeep WFPC2 data
in four optical filters \citep{W96,W00,C00}. The Hubble Deep Fields
(North and South) are specifically aimed at constraining cosmology and
galaxy evolution models, and in these studies  it is crucial to access
rest-frame optical wavelengths at high redshift through deep infrared
observations. Available ground-based NIR data from SOFI on the NTT
\citep{dC98} are not deep enough to match the space-based data. To 
fully take advantage of the deep optical data requires extremely 
deep wide-field imaging in the infrared at the best possible image 
quality; a combination that in the southern hemisphere can only be 
delivered by the \isaaclong\ \citep[ISAAC;][]{Mo97}, mounted on the 
Nasmyth-B focus of the 8.2 meter VLT Antu telescope. The infrared 
camera has a $2.5\arcmin \times 2.5 \arcmin$ field of view similar 
to that of the WFPC2 ($2.7\arcmin \times 2.7\arcmin$). 
ISAAC is equipped with a Rockwell Hawaii $1024 \times 1024$ HgCdTe 
array, offering imaging with a pixel scale of $0\farcs147$ pix$^{-1}$ 
in various broad and narrow band filters.
\par
\begin{figure*}[t]
\begin{center}
\includegraphics[width=0.7\textwidth]{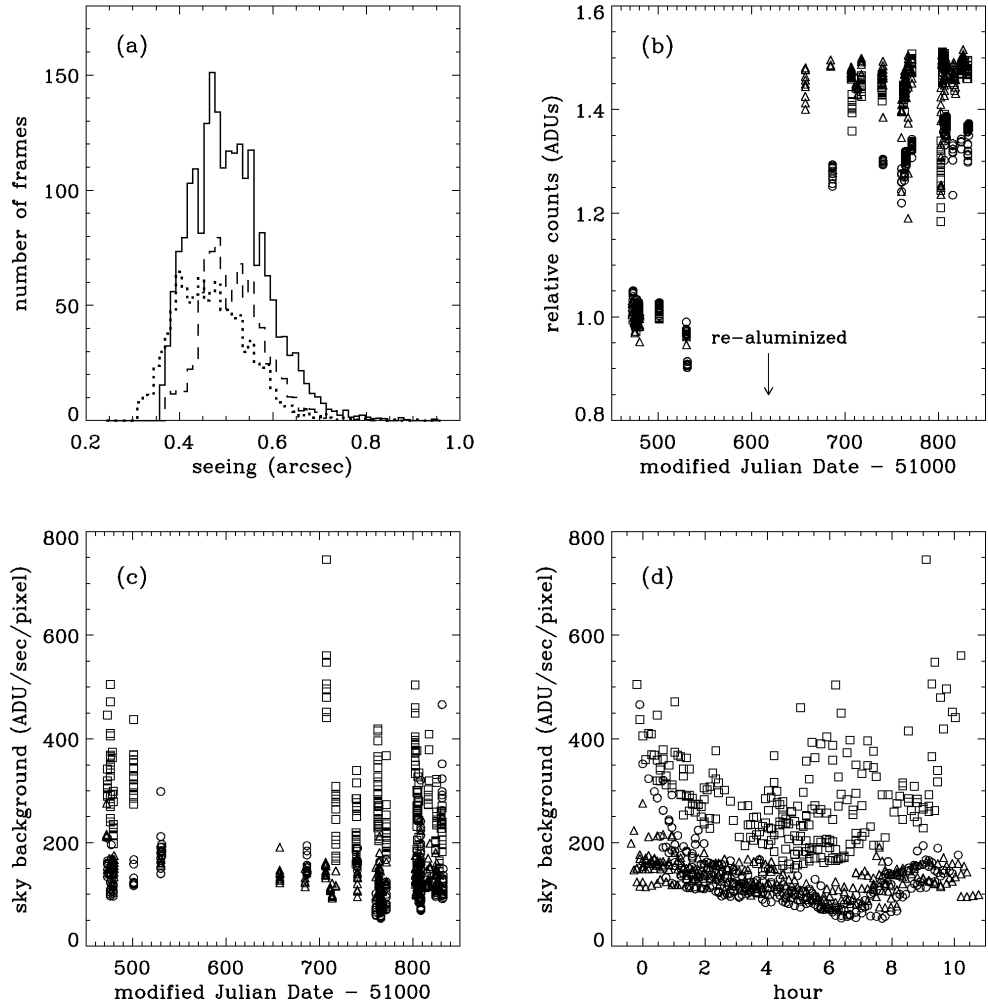}
\end{center}
\figcaption[Labbe.fig1.ps]
{  Shown are the raw data in the filters $J_s$ 
  ($dotted \ line$ or $circles$), $H$ ($dashed \ line$ or $squares$), 
  and $K_s$ ($solid \ line$ or $triangles$). {\it(a)} Histogram of the median
  seeing in the raw ISAAC images weighted by the weight function 
  of Eq. \ref{weight.ext} used to combine the images.
  (b) Relative instrumental counts in a
  $\approx3\arcsec$ radius aperture of four bright non-saturated stars
  in individual sky-subtracted exposures, plotted against Julian
  Date. The relative increase in counts, slightly dependent on
  wavelength, after cleaning and re-aluminization of the mirror
  directly reflects the increase in efficiency of the telescope,
  because the sky background levels {\it (c)} remained the
  same. Presumably, the photons were scattered by dirt rather 
  than absorbed before cleaning.
  {\it(d)} Nightly sky variations are largest and most rapid in
  the $H$-band and mean sky levels are highest at the beginning and
  ending of the night. $J_s$-band varies less and peaks at the start
  of the night, whereas $K$-band levels are most
  stable.\label{fig.obs}}
\end{figure*}

Our NIR imaging consists of a single ISAAC pointing centered on the
WFPC2 main-field  of the \hdfs\ ($\alpha = 22^h32^m55.464$, 
$\delta = -60\arcdeg 33 \arcmin 05.01\arcsec$, J2000)  
in the \vltj, \vlth\ and \vltk\ filters, which gives good sampling of
rest-frame optical wavelengths over the redshift range $1< z < 4$.  The \vltj\
filter is being established as the new standard broadband filter at
$\approx1.24\micron$ by most major observatories (Keck, Gemini,
Subaru, ESO), and is photometrically more accurate than the classical
$J$ because it is not cut off by atmospheric 
absorption. It is a top-hat filter with sharp edges, practically the same
effective wavelength as the normal $J$ filter, and half-transmittance
points at $1.17\micron$ and $1.33\micron$.
We used the \vltk\ filter which is bluer and narrower than standard
 $K$, but gives a better signal-to-noise ratio (SNR) for faint sources 
 because it is less affected by the high thermal background of the
  atmosphere and the telescope. The ISAAC \vlth\ and \vltk\ filters 
  are close to those used to establish the faint IR standard star 
  system \citep{P98}, while the \vltj\ filter requires a small color 
  correction. The WFPC2 filters that are used are  $F300W, F450W, F606W$ 
  and $F814W$ which we will call \wfu, \wfb, \wfv\ and \wfi, 
  respectively, where the subscript indicates the central wavelength 
  in nanometers.
\par

The observing strategy for the \hdfs\ follows established procedures for 
ground-based NIR imaging. The dominance of the sky background and its 
rapid variability in the infrared requires dithering of many short exposures. 
We used a $20\arcsec$ jitter box in which the telescope is moved in a 
random pattern of Poissonian offsets between successive exposures. This 
jitter size is a trade-off between keeping a large area at maximum depth 
and ensuring that each pixel has sufficient exposures on sky. Individual 
exposures have integration times of $6 \times 30~$s in \vltj, 
$6 \times 20~$s in \vlth, and $6 \times 10~$s in 
\vltk\ (subintegrations $\times $ detector integration times).  We 
requested service mode observations amounting to 32 hours in each 
band with a seeing requirement of $\lesssim 0\farcs5$, seeing 
conditions that are only available 25\% of the time at Paranal.
The observations were grouped in $112$  observation blocks (OBs), each of 
which uniquely defines a single observation of a target, including pointing, 
number of exposures in a sequence, and filter. The calibration plan for 
ISAAC provides the necessary calibration measurements for such blocks, 
including twilight flats, detector darks, and nightly zero points by 
observing LCO/Palomar NICMOS standard stars \citep{P98}.

\subsection{Observations}
\label{observations.observations}
The \hdfs\ was observed from October to December 1999 and from April
 to October 2000 under ESO program identification 164.O-0612(A). 
 A summary of the observations is shown in \ref{table.HDFS.summary}. 
 We obtained a total of $33.6$, $32.3$ and $35.6$ hours in
  \vltj, \vlth\ and \vltk, distributed over $33, 34$ and $55$ OBs, 
  or $1007, 968$ and $2136$ frames, respectively. This represents 
  all usable data, including aborted and re-executed OBs that were 
  outside weather specifications or seeing constraints. In the 
  reduction process these data are included with appropriate weighting 
  (see section \ref{reduction.additional}). Sixty-eight percent of 
  the data was obtained under photometric conditions and the 
  average airmass of all data was 1.25. A detailed summary of 
  observational parameters with pointing, observation date, image
   quality and photometric conditions can be found on the \fires\ 
   homepage on the World Wide Web (\fireswww).
\par

An analysis of various observational parameters reveals some
surprising trends in the data, whereas other expected relations are
less apparent. An overview is given in Figure \ref{fig.obs}. The
median seeing on the raw images is better than 0\farcs5 in all bands,
with the seeing of 90\% of the images in the range
$0\farcs4-0\farcs65$, as can be seen in Figure \ref{fig.obs}. 

Seeing may vary strongly on short timescales but it is not related to
any other parameter. The most drastic trend in the raw data is the
change of sensitivity with date. Since the cleaning and re-alumization
of the primary mirror in March 2000 the count rates of bright stars within
a $\approx 3\arcsec$ aperture increased by $+29$\% in \vltj, $+45$\%
in \vlth, and $+45$\% \vltk, which is reflected by a change in zero
points before and after this date. Because the average NIR sky levels
remained the same, this increase proportionally improved the
achievable signal-to-noise for background-limited sources. The change in
throughput was caused by light scattering, which explains why the sky
level remained constant. Sky levels
in \vltj\ and \vlth, dominated by airglow from OH-emission lines in
the upper atmosphere (typically 90 km altitude), vary unpredictably on
the timescale of minutes, but also systematically with observed
hour. The average sky level is highest at the beginning and end of
each night with peak-to-peak amplitudes of the variation being 50\%
relative to the average sky brightnesses over the night. The
background in $K_s$ is dominated by thermal emission of the telescope,
instrument, and atmosphere and is mainly a function of
temperature. The $K_s$ background is the most stable of all NIR bands
and only weakly correlated with airmass; our data do not show a strong
thermal atmospheric contribution, which should be proportional to
atmospheric path length. We take into account the variations of the
background and seeing through weighting in the data coadding process.

\section{Data Reduction}
The reduction process included the following steps: quality verification,
 flat-fielding, bad pixel correction, sky subtraction, distortion 
 correction, registration, photometric calibration and weighting of 
 individual frames, and combination into a single frame. We used a 
 modified version of the {DIMSUM}\footnote{DIMSUM is the Deep Infrared Mosaicing Software package developed by Peter Eisenhardt, Mark Dickinson, Adam Stanford, and John Ward, and is available via ftp to \url{ftp://iraf.noao.edu/contrib/dimsumV2/}} package and standard 
 routines in  {IRAF}\footnote{IRAF is distributed by the National Optical Astronomy Observatories, which are operated by the AURA, Inc., under cooperative agreement with the NSF.} 
 for sky subtraction and coadding, and the  {ECLIPSE}\footnote{ECLIPSE is an image processing package written by N. Devillard, and is available at \url{ftp://ftp.hq.eso.org/pub/eclipse/}} 
 package for creating the flatfields and the initial bad pixel masks. 
 We reduced the ISAAC observations several times with an increasing 
 level of sophistication, applying corrections to remove instrumental
  features, scattered light, or clear artifacts when required. Here
   we describe the first version of the reduction (v1.0) and the 
   last version (v3.0), leaving out the intermediate trial versions. 
   The last version produced the final \vltj, \vlth\ and \vltk\ images, 
   on which the photometry (see section \ref{sourcedetection}) and 
   analysis (see section \ref{analysis}) is based.

\subsection{Flatfields and Photometric Calibration}
\label{reduction.calib}

We constructed flatfields from images of the sky taken at dusk or
dawn, grouped per night and in the relevant filters, using the
\textbf{flat} routine in \textbf{ECLIPSE}, which also provided the bad
pixel maps. We excluded a few flats of poor quality and flats that
exhibited a large jump between the top row of the lower and bottom row
of the upper half of the array, possibly caused by the varying bias
levels of the Hawaii detector.  We averaged the remaining nightly
flats per month, and applied these to the individual frames of the OBs
taken in the same month. If no flatfield was available for a given
month we used an average flat of all months. The stability of these
monthly flats is very good and the structure changes little and in a
gradual way. We estimate the relative accuracy to be $0.2-0.4$\% per
pixel from the pixel-to-pixel rms variation between different monthly
flats. Large scale gradients in the monthly flats do not exceed 2\%.
We checked that standard stars, which were observed at various 
locations on the detector, were consistent within the error after 
flatfielding.
\par

Standard stars in the LCO/Palomar NICMOS list \citep{P98} were
observed each night, in a wide five-point jitter pattern. For each
star, on each night, and in each filter, we measured the instrumental
counts in a circular aperture of radius $20$ pixel ($2\farcs94$)  and
derived zero points per night from the magnitude of that star in the
NICMOS list. We identify non-photometric nights after comparison with
the median of the zero points over all nights before and after
re-aluminization in March 2000 (see section
\ref{observations.observations}). The photometric zero points exhibit
a large increase after March 2000 but, apart from this, the
night-to-night scatter is approximately 2\%. We adopted the mean of
the zero points after March 2000 as our reference value. See Table
\ref{table.ZP.summary} for the list of the adopted zero points. By
applying the nightly zero points to 4 bright unsaturated stars in the
HDF-S, observed on the same night under photometric conditions, we
obtain calibrated stellar magnitudes with a night-to-night rms
variation of only $\approx1-1.5$\%.
No corrections for atmospheric absorption were required because the
majority of the science data were obtained at similar airmass as the
standard star observations. In addition, instrumental count rates of
HDF-S stars in individual observation blocks reveal no correlation
with airmass. We used the calibrated magnitudes of the 4 reference
stars, averaged over all photometric nights, to calibrate every
individual exposure of the photometric and non-photometric
OBs. The detector non-linearity, as described by \citet{A01}, affects 
the photometric calibration by $\lesssim 1$\% in the $H$-band, 
where the exposure levels are highest. Because the effect is so small,
we do not correct for this.
We did not account for color terms due to differences between the
ISAAC and standard filter systems. \citet{A01} report that the ISAAC
\vlth\ and \vltk\ filters match very well those used to establish the
faint IR standard star system of \citet{P98}. Only the ISAAC \vltj\
filter is slightly redder than Persson's $J$ and this may introduce a
small color term, $\approx-0.04\cdot (J-K)_{LCO}$.  However, the
theoretical transformation between ISAAC magnitudes and those of
LCO/Palomar have never been experimentally verified. Furthermore, 
the predicted color correction is small and could not be 
reproduced with our data. In the absence of a better calibration we
chose not to apply any color correction. We did apply galactic 
extinction correction when deriving the photometric redshifts, 
see section \ref{phot.intro}, but it is not applied to the catalog.

\par
As a photometric sanity check, we compared $2\arcsec$ circular 
diameter aperture magnitudes of the brightest stars in the final 
(version 3.0, described below) images to magnitudes based on a 
small fraction of the data presented by R01. Each data set was 
independently reduced, the calibration based on different standard 
stars, and the shallower data were obtained before re-aluminization 
of the primary mirror. The magnitudes of the brightest sources in 
all bands agree within 1\% between the versions, indicating that
 the internal photometric systematics are well under control. For 
 the NIR data, the adopted transformations from the \citet{Jo66} 
 system to the AB system are taken from \citet{B88} and we apply 
 $J_{s,AB} =  J_{s,Vega} + 0.90, H_{AB} =  H_{Vega} + 1.38$ and 
 $K_{s, AB} = K_{s,Vega} + 1.86$. 

\subsection{Sky Subtraction and Cosmic Ray Removal}
The rapidly varying sky, typically 25 thousand times brighter than the 
sources we aim to detect, is the primary limiting factor in deep NIR imaging.
In the longest integrations, small errors in sky subtraction can
severely diminish the achievable depth and affect faint source
photometry. The {IRAF} package {DIMSUM} provides a two-pass routine to
optimally separate sky and astronomical signal in the dithered
images. We modified it to enable handling of large amounts of data and
replaced its co-adding subroutine, which assumes that the images are
undersampled, by the standard IRAF task
{IMAGES.IMMATCH.IMCOMBINE}. The following is a brief summary of the
steps performed by the {REDUCE} task in DIMSUM.

\par
For every science image in a given OB a sky image is constructed. After 
scaling the exposures to a common median, the sky is determined at each 
pixel position from a maximum of 8 and a minimum of 3 adjacent frames in
time. The lowest and highest values are rejected and the average of the
remainder is taken as the sky value. These values are subtracted from 
the scaled image to create a sky subtracted image. A set of stars is 
then used to compute relative shifts, and the images are integer 
registered and averaged to produce an intermediate image. All astronomical
 sources are identified and a corresponding object mask is created.
This mask is used in a second pass of sky subtraction where pixels 
covered by objects are excluded from the estimate of the sky. 
\par
The images show low-level pattern due to bias variations. Because they
generally reproduce they are automatically removed in the skysubtraction
step.
We find cosmic rays with DIMSUM, using a 
simple threshold algorithm and replacing them
by the local median, unless a pixel is found to have cosmic rays
in more than frames per OB. In this case the pixel is added to
the bad pixel map for that OB.

\subsection{First Version and Quality Verification}
The goal of the first reduction of the data set is to provide a 
non-optimized image, which we use to validate and to assess the 
improvements from more sophisticated image processing. The first 
version consists of registration on integer pixels and combination 
of the sky subtracted exposures per OB. For each of the $122$ OBs, 
we created an average and a median combined image to verify that 
cosmic rays and other outliers were removed correctly, and we visually 
inspected all $4149$ individual sky subtracted frames as well, finding 
that many required further processing as described in the following 
section. Finally, we generated the version 1.0 images (the first 
reduction of the full data set) by integer pixel shifting all OBs to 
a common reference frame, and coaveraging them into the  $J_s$, $H$ 
and $K_s$ images. While this first reduction is not optimal in terms 
of depth and image quality, it is robust owing to its straightforward 
reduction procedure.

\subsection{Additional Processing and Improvements}
\label{reduction.additional}
The individual sky subtracted frames are affected by a number of problems
 or instrumental features, which we briefly describe below, together with 
 the applied solutions and additional improvements that lead to the 
 version 3.0 images. The most important problems are:
\par
\begin{itemize}
\item Detector bias residuals, most pronounced at the rows where the read-out 
of the detector starts at the bottom (rows $1,2,...$) and halfway 
(rows $513, 514,...$), caused by the complex bias behaviour of the Rockwell 
Hawaii array. These variations are uniform along rows, and we removed the 
residual bias by subtracting the median along rows in individual sky subtracted 
exposures, after masking all sources.
\item Imperfect sky subtraction, caused by stray light or rapid background 
variations. Strong variations in the  backgrounds, reflection from high cirrus, 
reflected moonlight in the ISAAC optics or patterns of less obvious origin 
can lead to large scale residuals in the sky subtraction, particularly in 
\vltj\ and \vlth. For some OBs, we succesfully removed the residual 
patterns by splitting the sequence in two (in case of a sudden appearance 
of stray light), or subtracting a two-piece cubic spline fit along rows 
and columns to the background in individual frames, after masking 
all sources. We rejected a few frames, or masked the affected areas, 
if this simple solution did not work.
\item Unidentified cosmic rays or bad pixels. A small amount of bad pixels 
were not detected by ECLIPSE or DIMSUM routines but need to be identified 
because we average the final images without additional clipping or rejection. 
By combining the sky subtracted frames in a given OB without shifts and with 
the sources masked, we identified remaining cosmic rays or outliers through 
sigma clipping. We added $\sim60-100$ pixels per OB to the corresponding bad 
pixel map.
\end{itemize}
\par
Several steps were taken to improve the quality and limiting depths of the version 1.0 images, the most important of which are:
\begin{itemize}
 \item Distortion correction of the individual frames and direct 
 registration to the $3 \times 3$ blocked \wfi\ image 
 ($0.119\arcsec$pixel$^{-1}$), our preferred frame of reference. 
 We obtained the geometrical distortion coefficients for the 3rd 
 order polynomial solution from the ISAAC WWW-page\footnote{ISAAC home page: \url{ http://www.eso.org/instruments/isaac}}. The transformation procedure
  involves distortion correcting the ISAAC images, adjusting the
   frame-to-frame shifts, and finding the linear transformation 
  to the WFPC2 \wfi\ frame of reference. This linear transformation 
is the best fit mapping of 
source positions in the blocked WFPC2 \wfi\ image to the 
corresponding positions in the corrected \vltj-band 
image\footnote{We have noticed that the mapping solution changed slightly after the remount of ISAAC in March 2000, implying a 0.1\% scale difference.}. Compared to version 1.0 
described in the previous section, this procedure increases 
registration accuracy and image quality, decreases image smearing
at the edges introduced by the jittering and differential distortion.
Given the small amplitude of the ISAAC field distortions, the effect on 
photometry is negligible. In the linear transformation and distortion 
correction step the image is resampled once using a third-order 
polynomial interpolation, with a minimal effect of the interpolant 
on the noise properties. 

\item Weighting of the images. We substantially improved the final 
image depth and quality by assigning weights to individual frames 
that take into account changes in seeing, sky transparency, and 
background noise. Two schemes were applied: one that optimizes the 
signal-to-noise ratio (SNR) within an aperture of the size of the seeing disk, and one that optimizes the SNR per pixel. The first improves the detection efficiency of point sources, the other optimizes the surface brightness photometry. The weights $w_i$ of the frames are proportional to either the inverse scaled variance $zpscale_i \times var_i$ within a seeing disk of size  $s_i$, or to the inverse scaled variance per pixel, where the scaling $zpscale_i$ is the flux calibration applied to bring the instrumental counts of our four reference stars in the HDF-S to the calibrated magnitude.

\begin{equation}
  w_{i,point} \propto (zpscale_i \times var_i \times s^2_i)^{-1}
\end{equation}
\begin{equation}
 \label{weight.ext}
  w_{i,extended} \propto (zpscale_i \times var_i)^{-1}
\end{equation}
\end{itemize}

\subsection{Final Version and Post Processing}
The final combined \vltj, \vlth, and \vltk\ images (version 3.0) were 
constructed from the individually registered, distortion corrected, weighted 
and unclipped average of the $1007,\ 968$, and $2136$ NIR frames respectively. 
Ultimately, less than 3\% of individual frames were excluded in the final 
images because of poor quality. In this step we also generated the weight maps, 
which contain the weighted exposure time per pixel. We produced three versions
of the images, one with optimized weights for point sources, one with optimized
weights for surface brightness, and one consisting of the best quartile 
seeing fraction of all exposures, also optimized for point sources. The 
weighting has improved the image quality by 10--15\% and the background 
noise by 5--10\%, and distortion correction resulted in subpixel 
registration accuracy between the NIR images and \wfi-band image 
over the entire field of view.
\par
The sky subtraction routine in DIMSUM and our additional fitting of rows 
and columns (see section \ref{reduction.additional}) have introduced 
small negative biases in combined images, caused by systematic oversubtraction 
of the sky which was skewed by light of the faint extended PSF wings or very 
faint sources, undetectable in a single OB. Because of this, the flatness 
of the sky on large scales was limited to about 10$^{-5}$.  The negative 
bias was visible as clearly defined orthogonal stripes at P.A.$\approx6\arcdeg$,
as well as dark areas around the brightest stars or in the crowded parts of the 
images. To solve this, we rotated a copy of the final images back to the 
orientation in which we performed sky substraction, fitted a 3-piece cubic 
spline to the background along rows and columns (masking all sources), 
re-rotated the fit, and subtracted it. The sky in the final images is flat 
to a few $\times 10^{-6}$ on large ($>20\arcsec$) scales.

\section{Final Images}
\begin{figure*}[t]
\begin{center}
\includegraphics[width=0.9\textwidth]{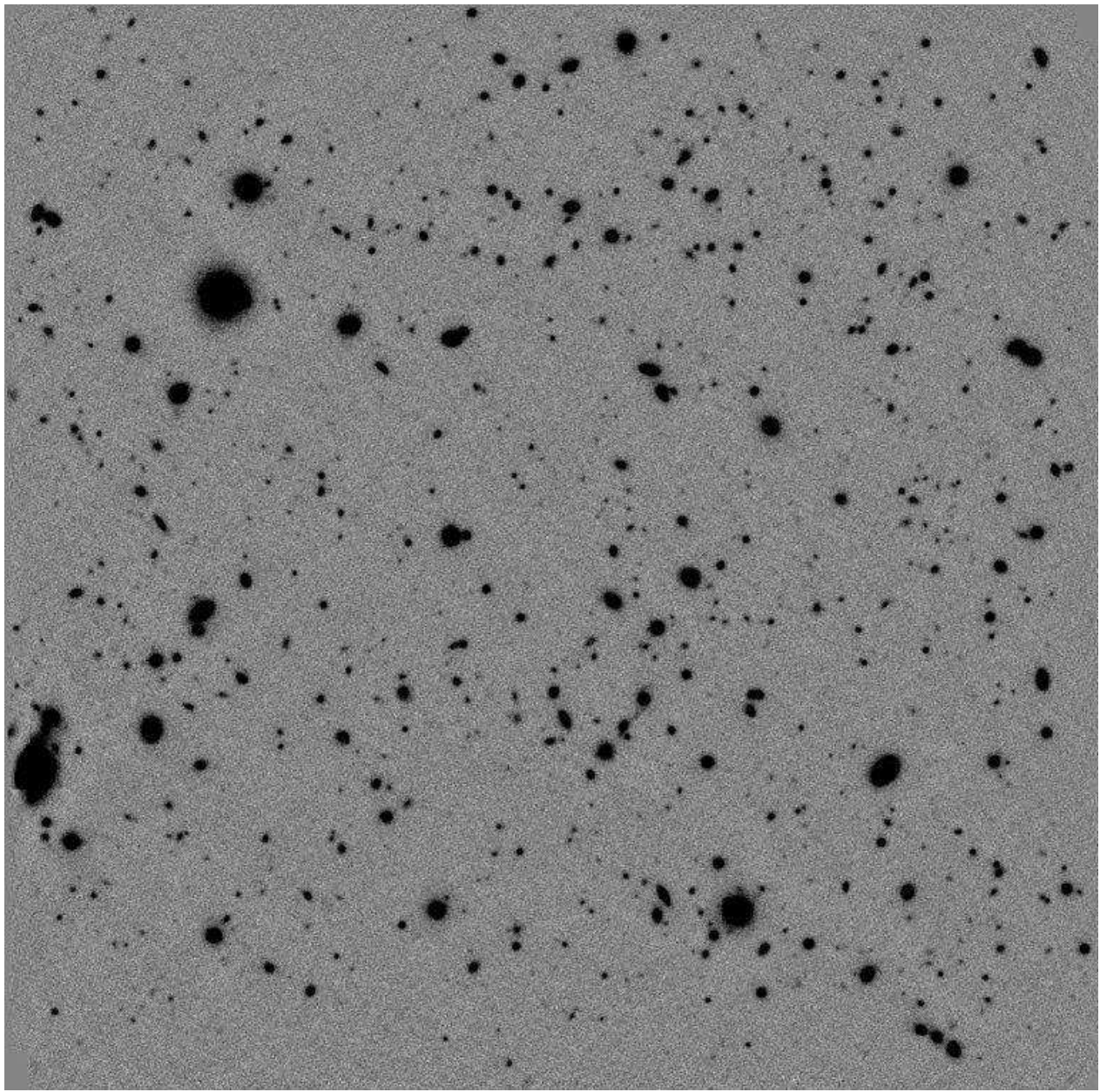} 
\end{center}
\figcaption[Labbe.fig2.ps]
{The HDF-S field in the ISAAC $K_s$-band divided
  by the square root of the weight map (based on the fractional exposure
  time per pixel) and displayed at linear scaling. The total
  integration time is 35.6 hours, the stellar FWHM$\approx0\farcs46$
  and the total field size is $2.85\arcmin \times
  2.85\arcmin$.\label{images.kdet}}
\end{figure*}

The reduced NIR \vltj, \vlth, and \vltk\ images and weight maps can be 
obtained from the \fires-WWW homepage  (\fireswww). Throughout the rest
 of the paper we will only consider the images optimized for point source 
 detection which we will use to assemble the catalog of sources. 

\subsection{Properties}
The pixel size in the NIR images equals that of the $3 \times 3$ blocked 
WFPC2 \wfi-band image at $0.119\arcsec$ pixel$^{-1}$. The combined ISAAC 
images are aligned with the HST version 2 images \citep{C00} with North 
up, and are normalized to instrumental counts per second. The images are 
shallower near the edges of the covered area because they received less 
exposure time in the dithering process, which is reflected in the weight 
map containing the fraction of total exposure time per pixel. The area of 
the ISAAC \vltk-band image with weight per pixel $w_K \geq 0.95, 0.2,$ 
and $0.01$ covers $4.5, 7.2$ and $8.3$ arcmin$^2$, while the area used 
for our preferred quality cut for photometry ($w \geq 0.2$ in all seven 
bands) is $4.7$ arcmin$^2$. The NIR images have been trimmed where the 
relative exposure time per pixel is less than 1\%.

\par
\begin{figure*}[t]
\begin{center}
\includegraphics[width=0.9\textwidth]{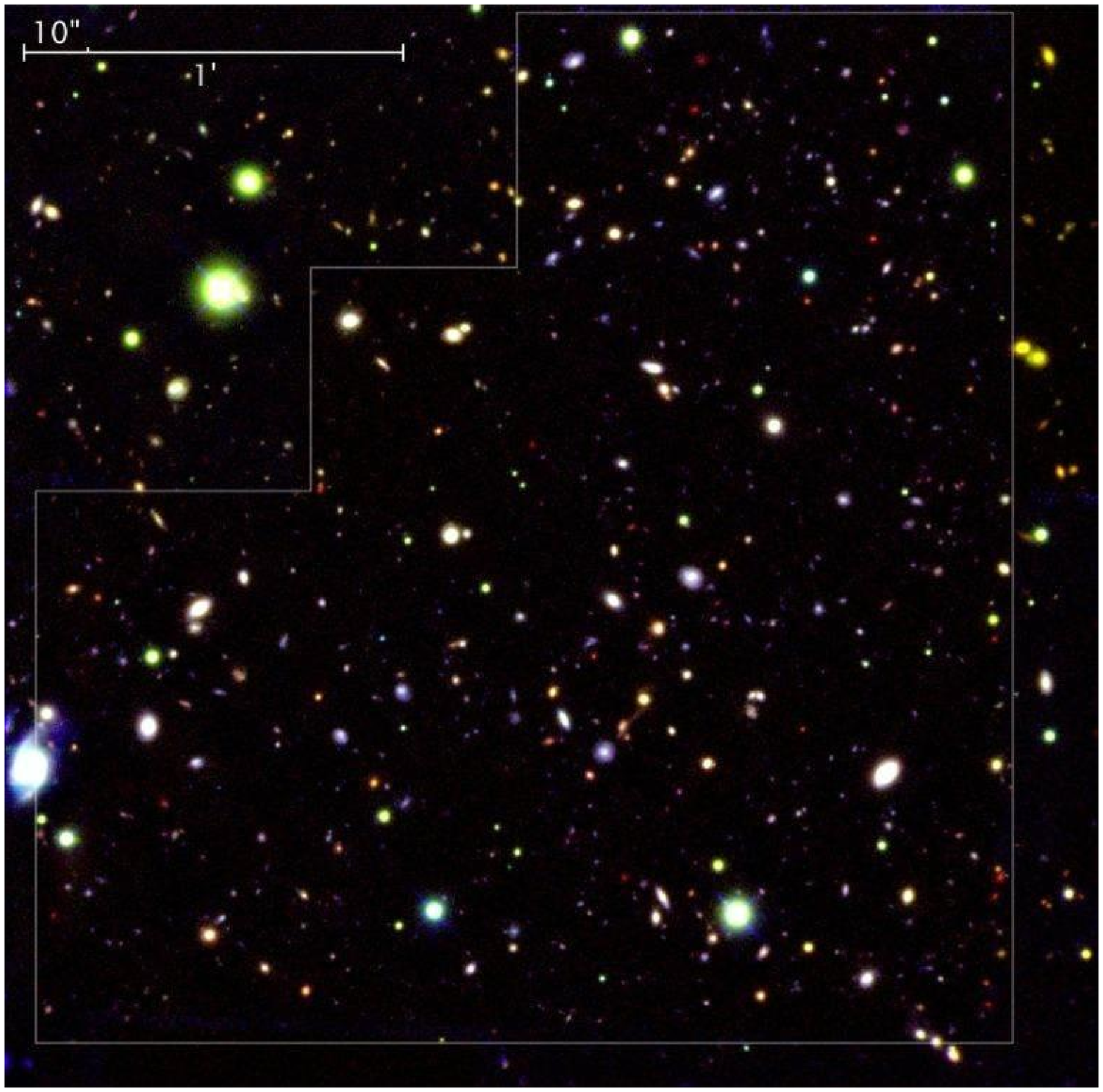}
\end{center}
\figcaption[Labbe.fig3.ps]
{Three-color composite image of the
  ISAAC field on top of the WFPC2 main-field and parts of three WFPC2
  flanking fields. The main-field is outlined in white and North is up. 
  The images are registered 
  and smoothed to a common seeing of FWHM$\approx 0\farcs46$, 
  coding WFPC2 $I_{814}$  in blue,
  ISAAC $J_s$ in green and ISAAC $K_s$ in red. There is a striking
  variety in optical-to-infrared colors, especially for fainter
  objects. A number of sources with red colors have photometric
  redshifts $z > 2$ and they are candidates for relatively 
  massive, evolved galaxies. These galaxies would not be 
  selected by the U-dropout technique because they are too faint 
  in the observer's optical. \label{images.ijk}}
\end{figure*}

\myfig\ \ref{images.kdet} shows the noise-equalized \vltk-detection 
image obtained by division with the square root of the exposure-time 
map. The richness in faint compact sources and the flatness of the 
background are readily visible.  \myfig\ \ref{images.ijk} shows 
a RGB color composite image of the  \wfi, \vltj, and \vltk\ images. 
The PSF of the space-based \wfi\ image has been matched to that 
of the NIR images at FWHM$\approx0\farcs46$ (see section \ref{images.quality}) 
and three adjacent WFPC2 \wfi\ flanking fields have been included for 
visual purposes. We set the linear stretch of both images to favor 
faint objects. Immediately striking is the rich variety in optical-NIR 
colors, even for the faint objects, indicating that the NIR observations 
are very deep and that there is a wide range of observed spectral 
shapes, which can result from different types of galaxies over a 
broad redshift range.

\subsection{Image Quality}
\label{images.quality}
The NIR PSF is stable and symmetric over the field with a gaussian 
core profile and an average ellipticity $< 0.05$ over the \vltj, 
\vlth, and \vltk\ images. The median FWHM of the profiles of ten selected 
isolated bright stars is $0\farcs45$ in \vltj, $0\farcs48$ \vlth, 
and $0\farcs46$ \vltk\, with $0\farcs04$ amplitude variation over 
the images. 
\par

For consistent photometry in all bands we convolved the measurement
images to a common PSF, corresponding to that of the \vlth-band which
had worst effective seeing (FWHM $= 0\farcs48$). The similarity of PSF
structure across the NIR images allowed simple gaussian smoothing for
a near perfect match. The complex PSF structure of the WFPC2 requires
convolving with a special kernel, which we constructed by deconvolving
an average image of bright isolated non-saturated stars in the
\vlth-band with the average \wfi-band image of the same stars. 
Division of the
stellar growth curves of the convolved images by the \vlth-band growth
curve shows that the fractional enclosed flux agrees to within 3\% at
radii $r \geq 0\farcs35$.

\begin{figure*} [t]
\begin{center} 
\includegraphics[width=0.55\textwidth]{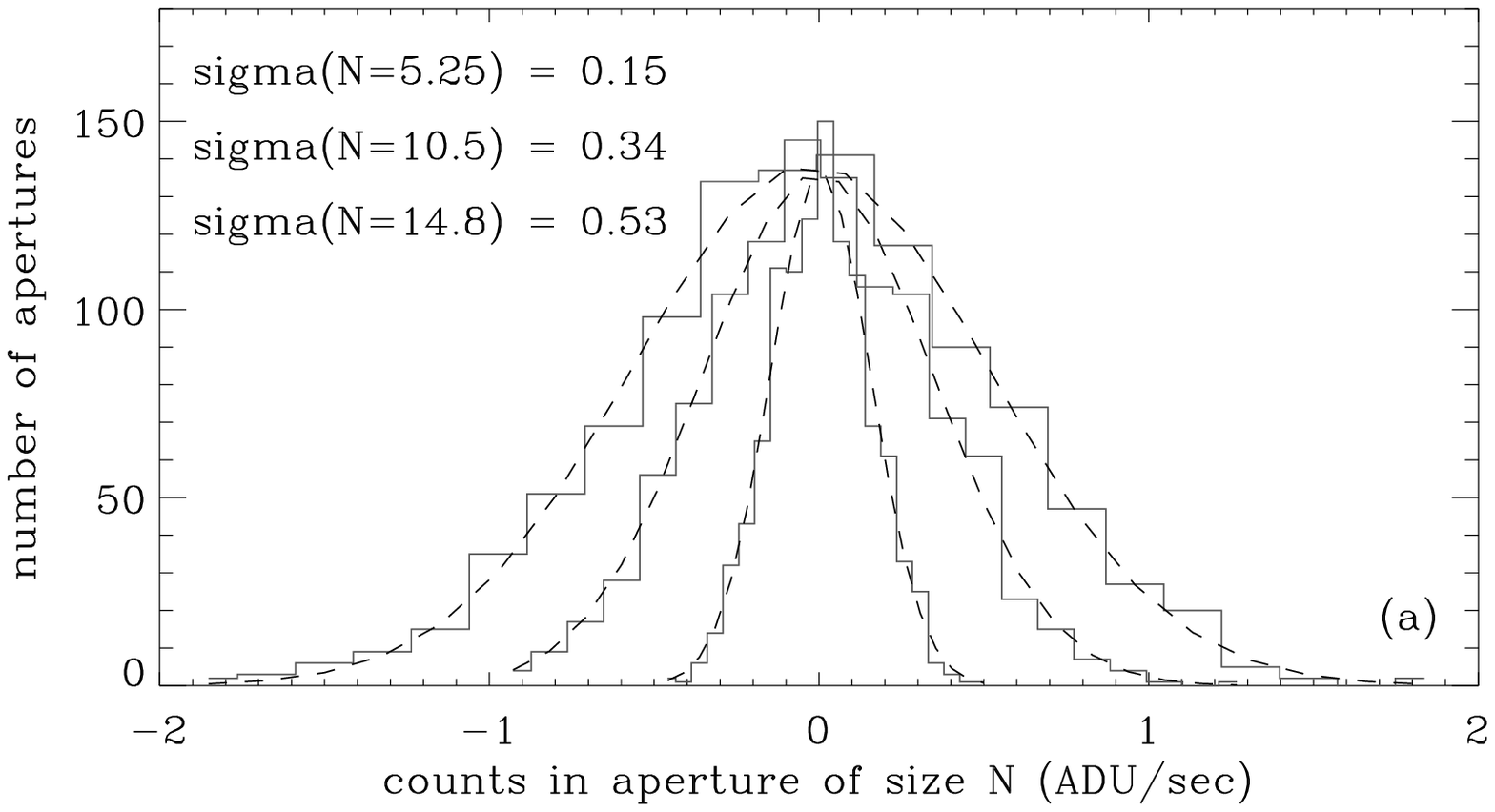}
\includegraphics[width=0.55\textwidth]{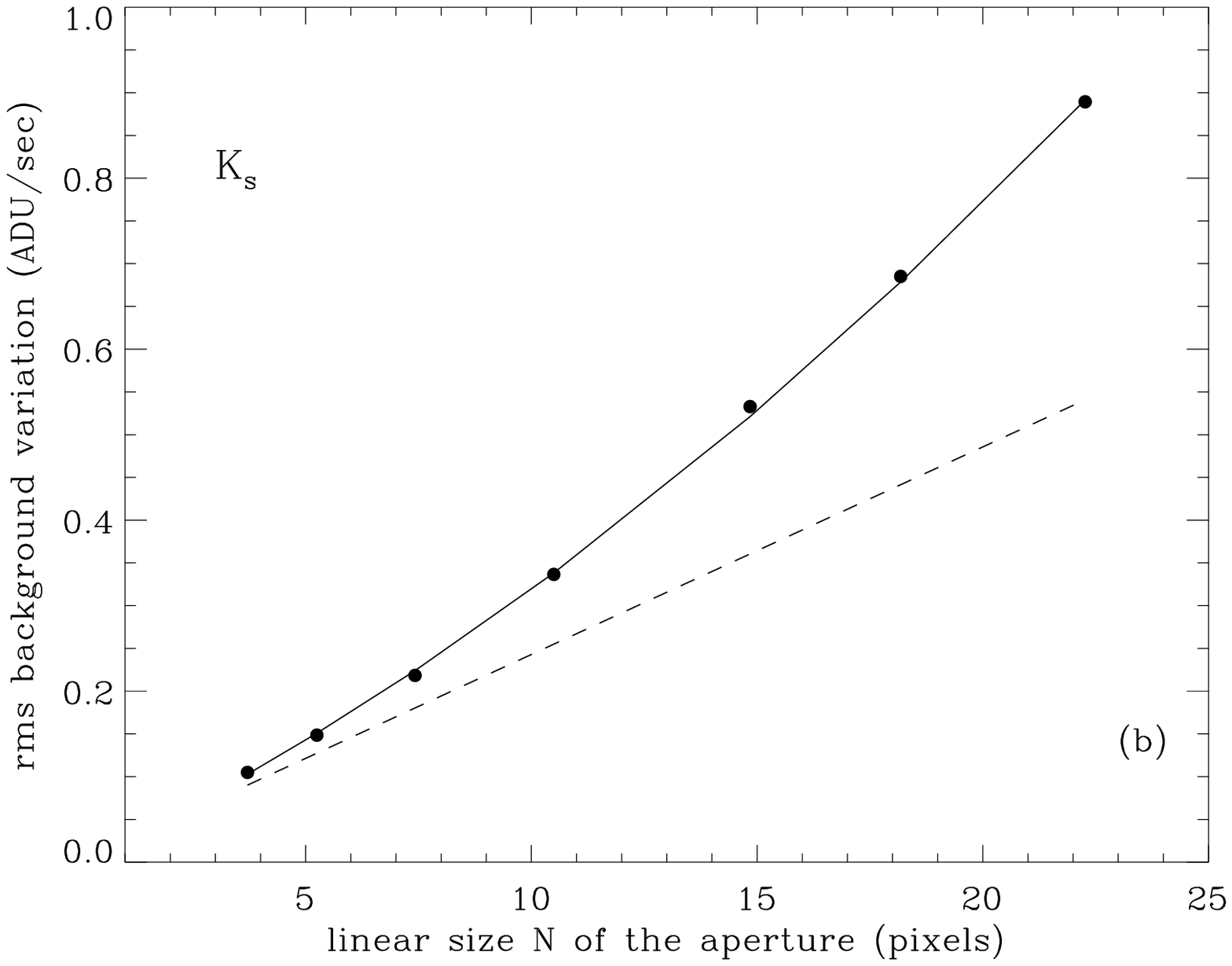}
\end{center}
\figcaption[Labbe.fig4.ps]
{Scaling relation of the measured
  background rms noise as a function of linear size $N=\sqrt{A}$ of
  apertures with area A. ({\it a}) Gaussians are fitted to histograms
  of $K_s$ counts in randomly placed apertures of increasing size, excluding
  pixels belonging to sources. This correctly accounts for
  pixel-to-pixel correlations and other effects, allowing us to
  measure the true rms variation as a function of linear size of
  aperture. ({\it b}) The $K_s$-band results ({\it solid points}),
  together with the best-fit scaling relation of
  Eq. \ref{error.scaling} ({\it solid line}), show that the measured variation
  in large apertures exceeds the variation expected from linear
  (Gaussian) scaling of the pixel-to-pixel noise ({\it dashed line}),
  likely due to  large scale correlated fluctuations of the
  background. \label{fig.errors}}
\end{figure*}

\subsection{Astrometry}

The relative registration between ISAAC and WFPC2 images needs to be
very precise, preferably a fraction of an original ISAAC pixel over
the whole field of view, to allow correct cross-identification of
sources, accurate color information and morphological comparison
between different bands. To verify our mapping of ISAAC to WFPC2
coordinates, we measured the positions  of the 20 brightest stars and
compact sources in all registered ISAAC exposures, and
we compared their positions with those in the \wfi\ image.  The rms
variation in position of individual sources is about $0.2-0.3$ pixel
at $0.119\arcsec$ pixel$^{-1}$ ($25-35$ mas), but for some sources
systematic offsets between the NIR and the optical up to $0.85$ pixel
($100$ mas) remain. The origin of the residuals is unclear and we
cannot fit them with low order polynomials. They could be real,
intrinsic to the sources, or due to systematic errors in the field
distortion correction of ISAAC or WFPC2\footnote{The ISAAC field
distortion might have changed over the years, but this cannot be
checked because recent distortion measurements are unavailable. The
worst case errors of relative positions across the four WFPC2 chips
can be $0\farcs1$ \citep{Vo97}, but is expected to be smaller for the
HDF-S images.}. However, for all our purposes, the
effect of positional errors of this amplitude is unimportant. The
error in absolute astrometry of the HST \hdfs\ coordinate system,
estimated to be less than $40$ mas, is dominated by the systematic
uncertainty in the positions of four reference stars \citep{C00,W00}.

\subsection{Backgrounds and Limiting Depths}
\label{images.background}

The noise properties of the raw individual ISAAC images
 are well described
by the variance of the signal collected in each pixel since both
Poisson and read noise are uncorrelated. However, image processing,
registration and combination have introduced correlations between 
neighbouring pixels and small errors in the background subtraction may also
contribute to the noise. Understanding the noise properties
well is crucial because limiting depths and photometric uncertainties
rely on them.

Instead of a formal description based on the analysis of the
covariance of correlated pixel pairs, we followed an empirical
approach where we fit the dependence of the rms background variation in
the image as a function of linear size $N = \sqrt{A}$ of apertures
with area $A$. Directly measuring the effective flux variations in
apertures of different sizes provides a more realistic estimate of signal
variations than formal Gaussian scaling $\sigma(N) = N\bar{\sigma}$ of
the pixel-to-pixel noise $\bar{\sigma}$, as is often done.

\par

We measured fluxes in $1200$  non-overlapping circular
apertures randomly placed on the registered convolved images, which were
also used for photometry. We excluded all
pixels belonging to sources detectable in \vltk\ at the 5$\sigma$
level (see section \ref{detection} for detection criteria). We used
identical aperture positions for each band $i$ and measured fluxes for
circular aperture diameters ranging from $0\farcs5$ to
$3\arcsec$.
 Then we obtained the flux dispersions by fitting a Gaussian
distribution to the histogram of fluxes at each aperture
size. Finally, we fitted a parameterized function of linear size to
the different dispersions:

\begin{equation}
\label{error.scaling}
  \sigma_i(N) =   N \bar{\sigma}_i (a_i + b_iN)/\sqrt{w_i}
\end{equation}

This equation describes the signal variation
versus aperture size $N$ over the entire image, taking into account
spatial variations as a result of relative weight $w_i$ for each passband $i$. 
As can be seen in \myfig\ \ref{fig.errors}, it provides a good fit
to the noise characteristics. The noise is significantly higher
than expected from uncorrelated (Gaussian) noise, indicated by a dashed line
in \myfig\ \ref{fig.errors}b.
\mytab\
\ref{table.background} shows the best fit values in all bands and the
corresponding limiting depths. The parameter $a$ reflects the
correlations of neighbouring pixels ($a > 1$), which is important in
the WFPC2 images because of heavy smoothing, but also in the ISAAC images 
given the resampling from $0\farcs147$ to $0\farcs119$ pixel$^{-1}$. 
The parameter $b$ accounts for large scale
correlated variations in the background ($b > 0$). This may be caused
by the presence of sources at very faint flux levels (confusion noise)
or instrumental features. Typically, the large scale correlated contribution
per pixel is only 3-15\% relative to the gaussian rms variation, 
but due to the $N^2$ proportionality the contribution to the variation 
in large apertures increases to significant levels. 
While the signal variations grow faster with area than
expected from a Gaussian, at any specific scale the variation is
consistent with a pure Gaussian. 

\par
From the analysis of the scaling relation of simulated colors we find
that part of the large scale irregularities in the background are
spatially correlated between bands. In particular, we measured the rms
variation of the \civ\ colors directly by subtracting 
in registered apertures the \wfi-band fluxes from the \wfv\ fluxes and fitting the
dispersion of the difference at each linear size. On large scales rms
variations are 30\% smaller than predicted from Eq. 3 if the noise
were uncorrelated. Yet, if we subtract the two fluxes in random
apertures, the scaling of the background variation is consistent with
the prediction. A similar effect is seen for the $I_{814}-J_s$ color,
but at a smaller amplitude. The spatial coherence of the background 
variations between filters and across cameras suggests that part of 
the background fluctuations may be associated with sources 
at very faint flux levels. Other contributions are likely similar 
flatfielding or skysubtraction residuals from one band to another.

\section{Source Detection and Photometry}
\label{sourcedetection}
The detection of sources at very faint magnitudes against a noisy background forces us to trade off completeness and reliability. A very low detection threshold may generate the most complete catalog, but we must then apply additional criteria to assess the reliability of each detection given that such a catalog will contain many spurious sources. More conservatively, we choose the lowest possible threshold for which contamination by noise is unimportant. We aim to produce a catalog with reliable colors suitable for robustly modeling of the intrinsic spectral energy distribution. Using SExtractor version 2.2.2 \citep{B96} with a detection procedure that optimizes sensitivity for point-like sources, we construct a \vltk-band selected catalog with seven band optical-to-infrared photometry.

\subsection{Detection}
\label{detection}
To detect objects with SExtractor using a constant signal-to-noise criterion 
over the entire image, including the shallower outer parts, we divide the point 
source optimized \vltk-image by the square root of the weight (exposure time)
 map to create a noise-equalized detection image. A source enters the catalog
  if, after low-pass filtering of the detection image, at least one pixel is 
  above $\approx5$ times the standard deviation of the filtered background, 
  corresponding to a total \vltk-band magnitude limit for point sources of 
  $K_s\approx26.0$. This depth is reached for the central $4.5$ arcmin$^2$.
   In total we have 833 detections in the entire survey area of $8.3$ 
   arcmin$^2$. Initially 820 sources are found but the detection software 
   fails to detect sources lying in the extended wings of the brightest 
   objects. To include these, we fit the surface brightness profiles of 
   the brightest sources with the GALPHOT package \citep{Fr89} in IRAF, 
   subtract the fit, and carry out a second detection pass with identical 
   parameters. Thirteen new objects enter the catalog, and 9 sources 
   detected in the first pass are replaced with improved photometry. 
   The catalog identification numbers of all second-pass objects start 
   at 10001, and the original entries of the updated sources are removed.

\par
Filtering affects only the detection process and the isophotal
parameters; other output parameters are affected only 
indirectly through barycenter and object extent. We chose a simple
two-dimensional gaussian detection filter (FWHM$=0\farcs46$),
approximating the core of the effective \vltk-band PSF well. Hence, we
optimize detectability for point-like sources, introducing a small
bias against faint extended objects. In principle it is possible to
combine multiple catalogs created with different filter sizes 
but merging these catalogs consistently is a complicated and 
subjective process yielding a modest gain only in sensitivity 
for larger objects. We prefer the small filter size equal to 
the PSF in the detection map because the majority of faint 
sources that we detect are compact or unresolved in the NIR and 
because we wish to minimize the blending effect of filtering on 
the isophotal parameters and on the confusion of sources. 
 SExtractor applies a {\it multi-thresholding} technique to 
 separate overlapping sources based on the distribution of the 
 filtered \vltk-band light. About 20\% of the sources are blended 
 because of the low value of the isophotal threshold; 
 in SExtractor this must always equal the detection threshold. 
 With the deblending parameters we used, the algorithm succeeds in 
 splitting close groups of separate galaxies, without ``oversplitting''
  galaxies with rich internal structure.

\par
We tested sensitivity to false detections by running SExtractor on a specially constructed \vltk-band noise map created by subtracting, in pairs, individual \vltk-band images of comparable seeing, after zero point scaling, and coaveraging the weighted difference images. This noise image has properties very similar to the noise in the original reduced image, including contributions from the detector and reduction process, but with no trace of astronomical sources. Our 
detection algorithm resulted in only 11 spurious sources over the full area.

\subsection{Optical and NIR Photometry}
\label{detection.photometry}
We use SExtractor's dual-image mode for spatially accurate and
consistent photometry, where objects are detected and isophotal
parameters are determined from the \vltk-band detection image while
the fluxes in all seven bands are measured in
the registered and PSF matched images. We used fluxes measured in circular
apertures $APER(D)$ with fixed-diameters $D$, isophotal apertures
$APER(ISO)$ determined by the $K_s$-band detection isophote at the $5\sigma$
detection threshold, and $APER(AUTO)$ (autoscaling) apertures inspired by
\citet{Kr80}, which scales an elliptical aperture based on the first
moments of the $K_s$-band light distribution. We select for each
object the best aperture based on simple criteria to enable detailed
control of photometry. We define two types of measurements:

\begin{itemize}
\item ``color'' flux, to obtain consistent and accurate colors. The optimal aperture is chosen based on the $K_s$ flux distribution, and this aperture is used to measure the flux in all other bands.
\item ``total'' flux, only in the \vltk-band, which gives the best estimate of the total $K_s$ flux. 
\end{itemize}
For both measurements we treat blended sources differently from unblended ones, 
and consider a source blended when its BLENDED {\em or} BIAS flag is set by SExtractor, 
as described in \citet{B96}.

\par
Our color aperture is chosen as follows, introducing the equivalent of a 
circular isophotal diameter $D_{iso} = 2\sqrt{A_{iso}/\pi}$ based on 
$A_{iso}$, the measured non-circular isophotal area within the detection-isophote:

\begin{equation}
  \begin{array}{lcl}
              \text{if the source is {\it not} blended}    & \;     &\; \\
         &\;  &\; \\     
           APER(COLOR) = \begin{cases}
                      APER(ISO) & (0\farcs7 < D_{iso} < 2\farcs0)\\
                      APER~(0\farcs7) & (D_{iso} \leq 0\farcs7)\\
                      APER~(2\farcs0) & (D_{iso} \geq 2\farcs0)\\
                     \end{cases}  &\;  &\; \\
              & \;     & \\
     \text{else if the source is blended}       &\;  &\; \\
         &\;  &\; \\     
       APER(COLOR)  = \begin{cases}
                      APER~(D_{iso}/s) & (0\farcs7 < D_{iso}/s < 2\farcs0)\\
                      APER~(0\farcs7) & (D_{iso}/s \leq 0\farcs7)\\
                      APER~(2\farcs0) & (D_{iso}/s \geq 2\farcs0)\\
                      \end{cases}  & \; &\; \\
  \end{array}
\end{equation}

\begin{figure*}[t]
\begin{center}
\includegraphics[width=0.6\textwidth]{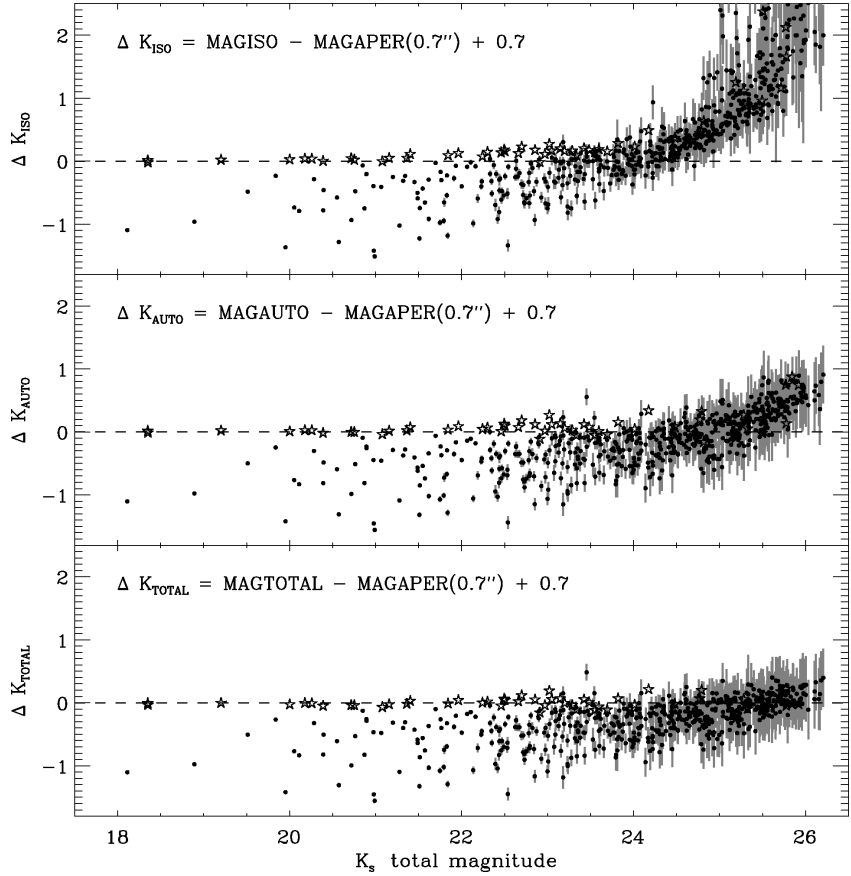}
\end{center}
\figcaption[Labbe.fig5.ps]
{Comparison of methods to estimate total $K_s$-band
  magnitude. Shown are isophotal ({\it top}), SExtractor's auto-scaling AUTO
  ({\it middle}), and our ``total'' magnitudes ({\it bottom}) as defined 
  in Eq. \ref{phot.total} and which are aperture corrected using 
  stellar growth curve analysis. 
  We subtracted the aperture corrected magnitude measured in an aperture
of $0\farcs7$ ($MAGAPER(0\farcs7) - 0.7$), 
 which produces the correct total magnitudes for stars and pointlike sources.
  Stars are marked by star symbols and fluxes are plotted
  with $\pm1\sigma$ error bars. The turn-up at $K_s \approx 24$ of
  isophotal and at $K_s \approx 25$ of the AUTO magnitudes shows that
  these photometric schemes systematically underestimate the total
  flux at faint levels, due to the decreasing size of the used aperture with
  magnitude. This effect is nearly absent in the bottom panel, which
  shows the total magnitudes measured in this paper. \label{fig.apercor}}
\end{figure*}

The parameter $s$ is the factor with which we shrink the circular apertures centered on blended sources, increasing the separation to the blended neighbour such that mutual flux contamination is minimal. This factor depends on the data set, and for our ISAAC $K_s$ image we find that $s=1.4$ is most successful. The smallest aperture considered, APER($0\farcs7$), $\approx 1.5$ FWHM of the effective PSF, optimizes the S/N for photometry of point sources in unweighted apertures and prevents smaller more error-prone apertures. The largest allowed aperture, APER($2\farcs0$), prevents large and inaccurate isophotal apertures driven by the filtered $K_s$-light distribution. We continuously assessed the robustness and quality of color flux measurements by inspecting the fits of redshifted galaxy templates to the flux points, as described in detail in $\S $\ref{phot.intro}. 
\par

We calculate the total flux in the $K_s$-band from the flux measured
in the AUTO aperture. 
 We define a circularized AUTO diameter $D_{auto} =
2\sqrt{A_{auto}/\pi}$ with $A_{auto}$ the area of the AUTO aperture,
and define the total magnitude as:

\begin{equation}
\label{phot.total}
  \begin{array}{lcl} 
        \text{if the source is {\it not} blended}  & \;   &\; \\
            &\;  &\; \\     
                APER(TOTAL) = APER(AUTO) \ \  \ \  &\;    &\; \\
            &\;  &\; \\     
         \text{else if the source is blended}    &\;  &\; \\
            &\;  &\; \\     
           APER(TOTAL) = APER(COLOR) \ \ \ \ &\;    &\;\\ 
     &\;  &\; \\
    \end{array}
\end{equation}

Finally, we apply an aperture correction using the growth curve of brighter stars 
to correct for the flux lost because it fell outside 
the ``total'' aperture. This aperture correction is necessary because 
it is substantial for our faintest sources, as shown in Figure 
\ref{fig.apercor} where we compare different methods to estimate
magnitude. The aperture correction reaches 0.7 mag 
at the faint end, therefore magnitudes are seriously underestimated if 
the aperture correction is ignored.

\par
We derive the $1\sigma$  photometric error for all measurements from
Eq. \ref{error.scaling} with the best-fit values shown in Table
\ref{table.background}. These errors may overestimate the uncertainty
in colors of adjacent bands (see section \ref{images.background}) but
it should represent well the photometric error over the entire $0.3\mu
- 2.2\micron$ wavelength range. The magnitudes may suffer from additional
uncertainties because of surface brightness biases or possible biases in 
the sky subtraction procedure which could depend on object magnitude and size.

\section{Photometric Redshifts}
\label{phot.intro}
To  physically interpret the seven-band photometry for our 
\vltk-band selected sample, we use a photometric redshift 
($z_{phot}$) technique explained in detail by R01. In summary,
 we correct the observed flux points for galactic extinction
 (see \citealt{SFD98}) and we model the rest-frame colors 
 of the galaxies by fitting a linear combination of 
 redshifted empirical galaxy templates. The redshift with 
 the lowest $\chi^2$ statistic, where
\begin{equation}
  \chi^2(z) = \sum^{N_{filter}}_{i=1} \left[ \frac{F_i^{data}-F_i^{model}}{\sigma_{i}^{data}} \right]^2
\end{equation}
 is then chosen as the most likely $z_{phot}$. Using a 
 linear combination of SEDs as $F^{model}$ minimizes the 
 a priori assumptions about the nature and stellar composition 
 of the detected sources.

Our data set with three deep NIR bands samples the position 
of Balmer/4000\AA\ break over $1 \lesssim z \lesssim 4$, allowing 
us to probe the redshift distribution of more evolved galaxy 
types that may have little rest-frame UV flux and hence a weak 
or virtually absent Lyman break.

\begin{figure*}[t]
\begin{center}
\includegraphics[width=0.7\textwidth]{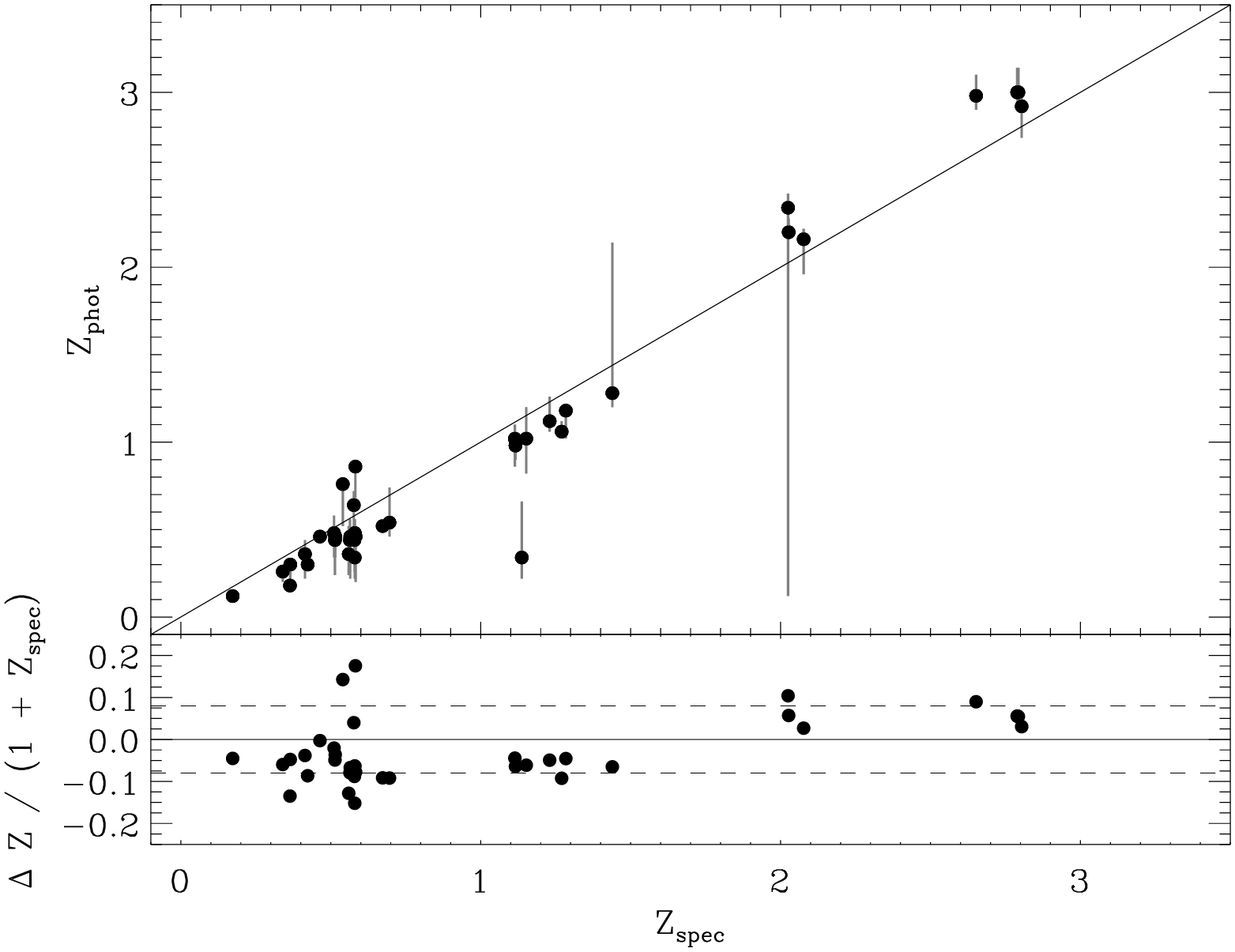}
\end{center}
\figcaption[Labbe.fig6.ps]
{Direct comparison of photometric
  redshifts to the 39 spectroscopic redshifts of objects in the HDF-S
  with good photometry in all bands. The 68\% error bars are derived
  from our Monte Carlo simulations and the diagonal line corresponds
  to a one-to-one relation to guide the eye. While the agreement is
  excellent with no failures for this small sample and with mean
  $\Delta z/(1 + z) = 0.08$, large asymmetric uncertainties remain for
  some objects indicating the presence of a second photometric redshift
  solution of comparable likelihood at a  different
  redshift. \label{fig.zphot_zspec}}
\end{figure*}
 
\subsection{Photometric Templates}
We used the local Hubble type templates E, Sbc, Scd, and Im from
\citet{CWW80}, the two starburst templates, SB1 and SB2, with low
derived reddening from \citet{K96}, and a 10 Myr old single age
template model from \citet{BC01}. The starburst templates are needed
because many galaxies even in the nearby Universe have bluer colors
than the bluest CWW templates. The observed templates are extended
beyond their published wavelengths into the far-ultraviolet by power
law extrapolation and into the NIR using stellar population synthesis
models from \citet{BC01}, with the initial mass functions and
star-formation timescales  for each template Hubble type from
\citet{Po96}. We accounted for internal hydrogen absorption of each
galaxy by setting the flux blueward of the $912$\AA\ Lyman limit to
zero, and for the redshift-dependent cosmic mean opacity due to
neutral intergalactic hydrogen by following the prescriptions of \citet{M95}.

\subsection{$Z_{phot}$ Uncertainties}
\label{photz.uncertainty}
The best test of photometric redshifts is direct comparison to
spectroscopic redshifts, but spectroscopic redshifts in the HDF-S are
still scarce. We calculate the uncertainty in the photometric redshift
due to the flux measurement errors using a Monte-Carlo (MC) technique
derived from that used in R01 and fully explained in \citet{Ru02b}. At
bright magnitudes template mismatch dominates the errors, something
that is not modeled by the MC simulation.
Hence, the MC error bars for bright galaxies are severe
underestimates. At fainter magnitudes, the uncertainty is driven by
errors in photometry \citep{FLY99} and the MC technique should provide
accurate $z_{phot}$ uncertainties. Experience from R01 showed that two
ways to correct for the template mismatch, setting a minimum fractional flux
error or setting a minimum $z_{phot}$ error based on the mean disagreement
with $z_{spec}$, either degrade the accuracy
of the $z_{phot}$ measurement or reflect the systematic error only in
the mean, while template mismatch can be a strong function of SED
shape and redshift. A method based completely on Monte-Carlo
techniques is preferable because it has a straightforwardly computable
redshift probability function. This approach is desirable for
estimating the rest-frame luminosities and colors \citep{Ru02b}.
\par

Therefore, we modify the MC errors directly using the FIRES
photometry. 
In summary, we estimate the systematic component of the $z_{phot}$
uncertainty by scaling up all the photometric errors for a given galaxy
with a constant to bring residuals of the fit in
agreement with the errors.
This will not change the best fit redshift
and SED and will not modify the MC errorbars of faint objects, but it
will enlarge the redshift interval over which the templates can
satisfactorily fit the bright objects. Only in case of widely
different photometric errors between the visible and infrared might
 the modified MC uncertainties still underestimate the true $z_{phot}$
uncertainty.

\par
In Figure \ref{fig.zphot_zspec} we show a direct test of photometric
 redshifts of the 39 objects in the HDF-S with available spectroscopy 
 and good photometry in all bands. The current set of spectroscopic 
 redshifts in the \hdfs\ will appear in \citet{Ru02a}. For the small 
 sample that we can directly compare, we find excellent agreement 
 with no failures and with a mean $\Delta z/(1 + z_{spec}) \approx 0.08$ 
 with $\Delta z = |z_{spec} - z_{phot}|$. It is encouraging to see that 
 the modified 68\% error bars that were derived from the Monte Carlo 
 simulations are consistent with the measured disagreement between 
 $z_{phot}$ and the $z_{spec}$ in the \hdfs. However, large asymmetric 
 uncertainties remain for some objects, clearly showing the presence 
 of a second photometric redshift solution of comparable likelihood 
 at a vastly different redshift, revealing limits on the applicability 
 of the photometric redshift technique.

\begin{figure*}[t]
\begin{center}
\includegraphics[width=0.7\textwidth]{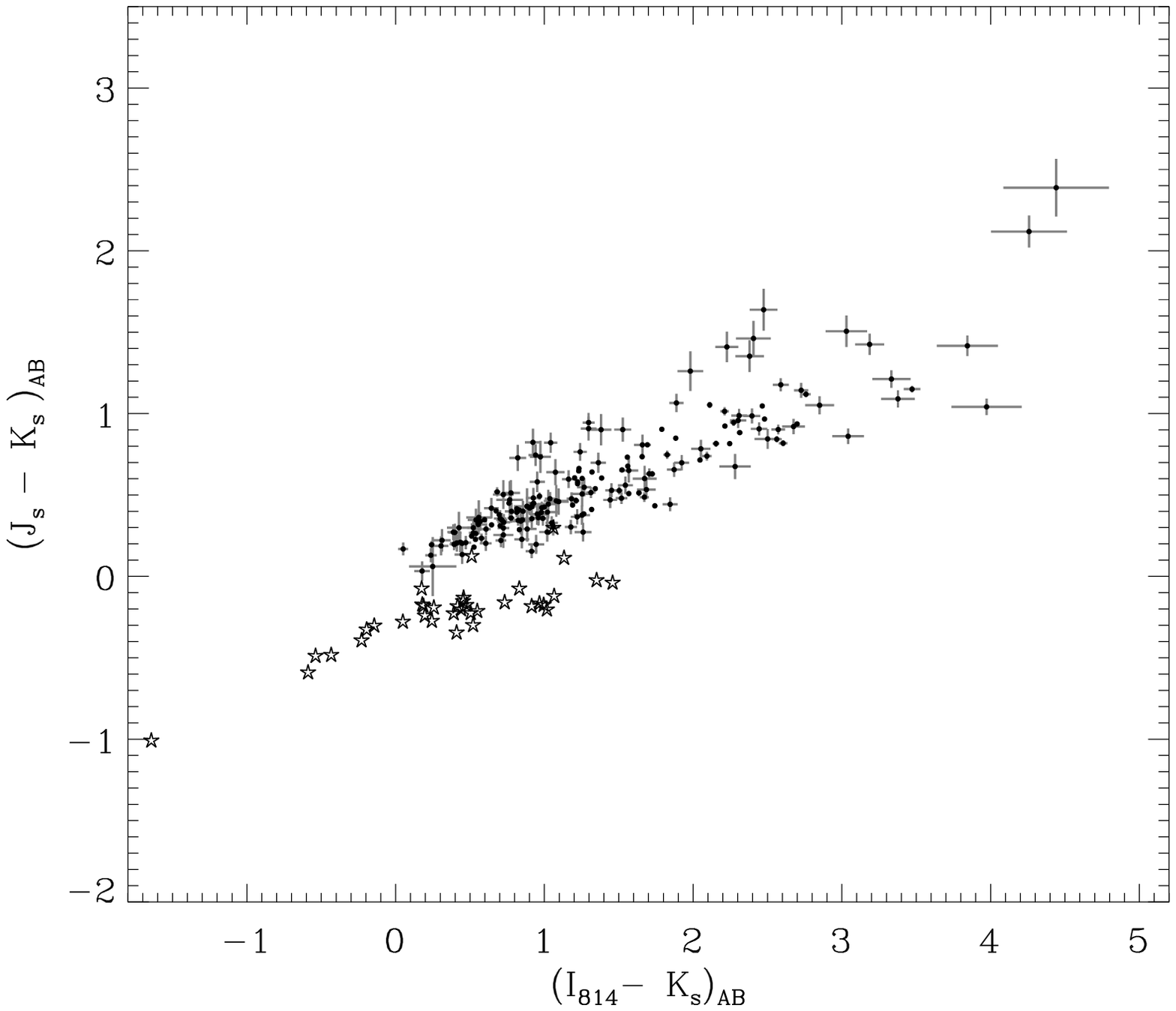}
\end{center}
\figcaption[Labbe.fig7.ps]
{\cjk\ versus \cik\
color-color diagram (on the AB system) for the sources with $K_s < 24$
in the HDF-S with a minimum of 20\% of the total exposure time in all
bands. Identified stars are marked by a star symbol.  The colors are
plotted with $\pm1\sigma$ error bars. There is a large variation in
both $I - K_s$ and $J_s - K_s$ colors.  Redshifted galaxies are well
separated from the stellar locus in color-color
space. \label{fig.cc_IK_JK}}
\end{figure*}

\subsection{Stars}
\label{photz.stars}
In a pencil beam survey at high Galactic latitude such as the \hdfs, a 
limited number of foreground stars are expected. We identify stars as 
those objects which have a better raw $\chi^2$ for a single stellar 
template fit than the $\chi^2$ for the galaxy template combination. 
The stellar templates are the NEXTGEN model atmospheres from \citet{Ha99} 
for main sequence stars with temperatures of $3000$ to $10000$K, 
assuming local thermodynamical equilibrium (LTE). Models of cooler 
and hotter stars cannot be included because non-LTE effects are 
important. We checked the resulting list of stars using the FWHM in 
the original \wfb-band image and the \cjk\ color, excluding two objects 
(catalog IDs 207 and 296) that were obviously extended in \wfb, and 
we find a total of \starcounts\ stars. As shown in Figure \ref{fig.cc_IK_JK}, 
most galaxies are clearly separated from the stellar locus in $I_{814} - K_s$ 
versus $J_s - K_s$ color-color space.
Other cooler stars might still resemble SEDs of redshifted compact galaxies 
but the latter are generally redder in the infrared \cjk\ than most known 
M or methane dwarfs. Known cool L-dwarfs fall along a redder extension 
of the track traced by M-dwarfs in color-color space and have progressively
 redder $J_s - K_s$ colors for later spectral types. However measurements 
 by \citet{Ki00} show the L-dwarf sequence abruptly stopping at 
 $(J_s - K_s)_J \approx 2.1$ (the subscript noting Johnson magnitudes, 
 see section \ref{reduction.calib} for the transformations to the AB system) 
 whereas even cooler T-dwarfs have much bluer $(J_s - K_s)_J\approx 0$ 
 colors than expected from their temperatures due to strong molecular 
 absorption. This is important because if we would apply a $(J_s - K_s)_J > 2.3$ 
 photometric criterion to select $z > 2$ galaxies (as discussed in section 
 \ref{results.colors}), then we should ensure that cool Galactic stars are 
 not expected in such a sample. The published data on the lowest-mass stars 
 suggest that they are too blue in infrared colors to be selected this way. 
 Only heavily reddened stars with thick circumstellar dust shells, such as 
 extreme carbon stars or Mira variables, or extremely metal-free stars 
 having a hypothetical $\lesssim 1500$K blackbody spectrum  could also 
 have red $(J_s - K_s)_J > 2.3$ colors but it seems unlikely that the tiny 
 field of the \hdfs\ would contain such unusual sources.

\begin{figure*} [t]
\begin{center}
\includegraphics[width=0.6\textwidth]{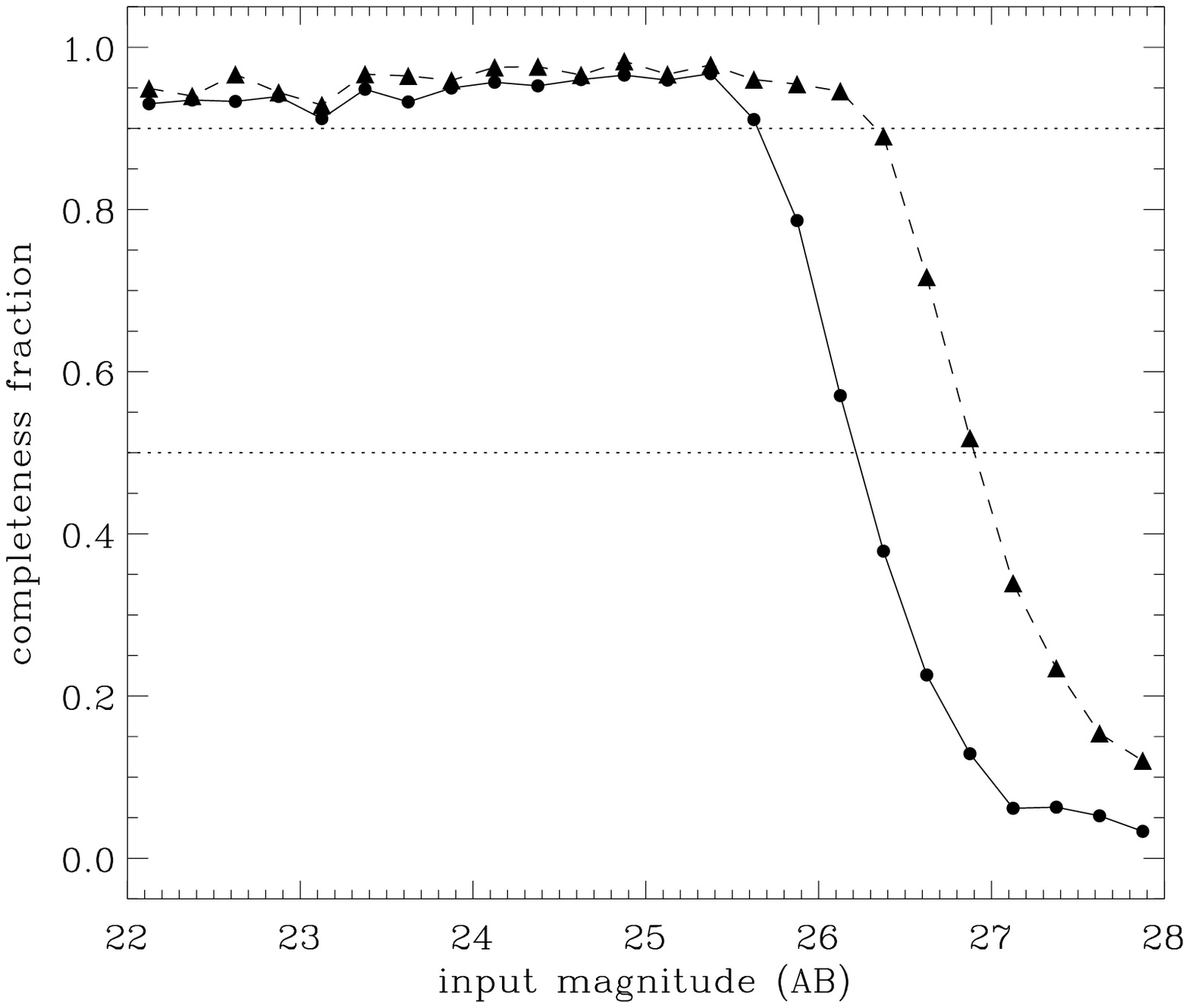}
\end{center}
\figcaption[Labbe.fig8.ps]
{Completeness curves (on the AB
system) for the detectability of point sources in $J_s$ ({\it
triangles}) and $K_s$ ({\it points}), based on simulations where we
calculated the recovered fraction of stars that were dimmed to
magnitudes between $22$ and $28$ and embedded in the survey
images. The detection threshold of the source extraction software was
set to $3.5\sigma$ of the filtered background rms. The dotted lines
indicate the 50\% and 90\% completeness
levels. \label{fig.complete_jk} }
\end{figure*}

\section{Catalog Parameters}
The \vltk-selected catalog of sources is published electronically. We
describe here a subset of the photometry containing the most important parameters.
The catalog with full photometry and explanation can be obtained from the 
FIRES homepage \footnote{\fireswww}.  

\begin{itemize}
\item { $ID$}. --- A running identification number in catalog order as 
reported by SExtractor. Sources added in the second detection pass have 
numbers higher than 10000.
\item { $x,y$}. --- The pixel positions of the objects corresponding to 
the coordinate system of the original (unblocked) WFPC2 version 2 images.

\item { $RA, DEC$}. -- The right ascension and declination in equinox
  J2000.0 coordinates of which only the minutes and seconds of right
  ascension, and negative arcminutes and arcseconds of declination are
  given. To these must be added $22^h$(R.A.) and $-60\arcdeg$(DEC).

\item { $f_{col,i} \pm \sigma_i$}. --- The sum of counts in the ``color'' aperture $f_{col,i}$ in band $i=\{ U_{300},B_{450},V_{606},I_{814},J_s,H,K_s \}$ and its simulated uncertainty $\sigma_i$, as described in $\S$ \ref{detection.photometry}.
The fluxes are given in units of $10^{-31}$ ergs s$^{-1}$ Hz$^{-1}$ cm$^{-2}$. 

\item { $K_{tot} \pm \sigma (K_{tot})$}.  --- Estimate of the total $K_s$-band flux and its uncertainty. The sum of counts in the ``total'' aperture is corrected for missing flux assuming a PSF profile outside the aperture, as described in $\S$ \ref{detection.photometry}.

\item { $ap\_col$}. ---  An integer encoding the aperture type that was used to measure {$f_{col,i}$}. This is either a (1) $0\farcs7$  diameter circular aperture, (2) $2\farcs0$  diameter circular aperture, (3) isophotal aperture determined by the
detection-image isophote, or a (4) circular aperture with a reduced isophotal diameter $D = \sqrt{(A_{iso}/\pi)}/1.4$.
\item { $ap\_tot$}. --- An integer encoding the aperture type that was used to measure $K_{tot}$. This is either a (1) automatic Kron-like aperture, or a (2) circular aperture within a reduced isophotal diameter.  
\item { $r_{col}, r_{tot}$}. --- Circularized radii $r = \sqrt{A/\pi}$, corresponding to the area $A$ of the specified ``color'' or ``total'' aperture. 
\item { $A_{iso}, A_{auto}$}. --- Area of the detection isophote $A_{iso}$ and area of the autoscaling elliptical aperture $A_{auto} = \pi*a*b$ with semi-major axis $a$ and semi-minor axis $b$.
\item { $FWHM_K, FWHM_I$}. --- Full width at half maximum of a source in the \vltk\ detection image $FWHM_K$, and that of the brightest \wfi-band source that lies in its detection isophote $FWHM_I$. We obtained the latter by running SExtractor separately on the original \wfi-image and cross correlating the \wfi-selected catalog with the \vltk-limited catalog. 
\item { $w_i$}. --- The weight $w_i$ represents, for each band $i$, the
  fraction  of the total exposure time at the location of a source.
\item { $flags$}.--- Three binary flags are given. The $bias$ flag indicates either that the AUTO aperture measument is affected by nearby sources, or marks apertures containing more than 10\% bad pixels. The $blended$ flag indicates overlapping sources, while the $star$ flag shows that the source SED is best fit with a stellar template (see section \ref{photz.stars}).
\end{itemize}

\begin{figure*}[t]
\begin{center}
\includegraphics[width=0.6\textwidth]{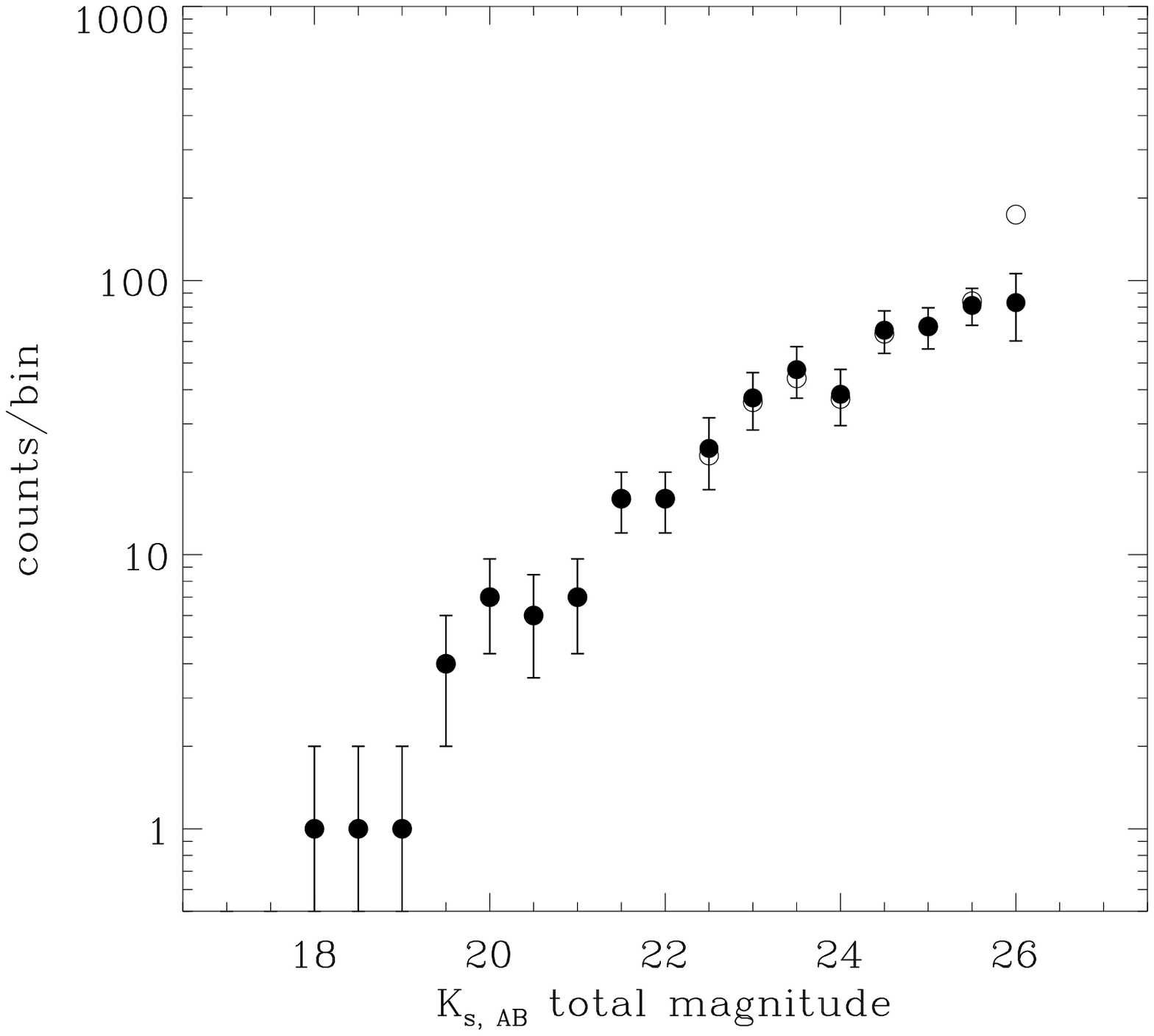}
\end{center}
\figcaption[Labbe.fig9.ps]
{Differential $K_s$-band
counts (on the AB system) of galaxies in the \hdfs. The counts are
based on auto-scaling apertures \citep{Kr80} for isolated sources and
adapted isophotal apertures for blended sources, both corrected to total
magnitudes using stellar growth curve measurements. Raw counts ({\it
open circles}) and counts corrected for incompleteness and false
positive detections using point source simulations ({\it filled
circles}) are shown. The small corrections at magnitudes $\gtrsim 23$
reflect missed sources due to confusion. Effective corrections at the
faintest magnitudes $K_s \sim 25 - 25.5$ are very small because the
loss of sources on negative noise regions (incompletenesss) is
compensated by the number of sources pushed above the detection limit
by positive noise fluctuations. Only the faintest $0.5$ mag bin
centered on $K_s=26.0$, bordering the $3.5\sigma$ detection limit
($K_s\approx26.3$), is significantly corrected because of a
contribution of false positive detections.\label{fig.counts_k}}
\end{figure*}

\section{Analysis}
\label{analysis} 
\subsection{Completeness and Number Counts}
\label{detection.completeness}
The completeness curves for point-sources in the \vltj\ and \vltk-band 
as a function of input magnitude are shown in \myfig\ \ref{fig.complete_jk}.  
Our 90\% and 50\% completeness levels on the AB magnitude system are $25.65$ 
and $26.25$, respectively, in \vltk, and $26.30$ and $26.90$ in \vltj.  

\par
We derived the limits from simulations where we extracted a bright 
non-saturated star from the survey image and add it back $30000$ times 
at random locations, applying a random flux scaling drawn from a 
rising count slope (or an increasing surface density of 
galaxies with magnitude) to bring it to magnitudes between 
$22 \leq K_{s,AB} \leq 28$. We added the dimmed stars back in series 
of $30$ realizations so that they do not overlap each other. 
The rising count slope needs to be considered because the slope 
influences the  number of recovered galaxies per apparent 
magnitude, as described below.  The input count slope is 
based on the observed surface densities in the faint magnitude range where 
the signal-to-noise is $60 \lesssim SNR \lesssim 10$ 
(or $23 \lesssim K_{s,tot} \lesssim 25$) and where incompleteness does not yet play a role. 
We used only the deepest central $4.5$ arcmin$^2$ of the  \vltj\  and \vltk\ 
images ($w > 0.95$) with near uniform image quality and exclude four small 
regions around the brightest stars. In the simulation images we extract 
sources following the same procedures as described in $\S$\ref{detection}, 
but applying a reduced ($\approx 3.5\sigma$) detection threshold. We measure 
the recovered fraction of input sources against apparent magnitude, and from 
this we estimate the detection efficiency of point-like sources which we 
use to correct the observed number counts. We executed this procedure 
in the \vltj\ and \vltk-band. 

\par
The resulting completeness curves assume that the true profile of the 
source is point-like and therefore they should be considered upper limits. 
An extended source would have brighter completeness limits depending on 
the true source size, its flux profile, and the filter that is used 
in the detection process. However, the detailed treatment of detection 
efficiency as a function of source morphology and detection criteria 
is beyond the scope of this paper. 
 \par
When using the completeness simulations to correct the number counts, 
we choose a simple approach and apply a single correction down to 
$\approx 50$\% completeness, based on the ratio of the simulated 
counts per input magnitude bin to the total recovered counts per 
observed total magnitude bin. More sophisticated modeling is possible 
but requires detailed knowledge of the intrinsic size and shape distribution 
of faint NIR galaxies. The simple approach corrects for all effects 
resulting from detection criteria, photometric scheme, incompleteness, 
and noise peaks. 
We find it works well if the total magnitude of sources is measured 
correctly, with little systematic difference between the input and 
recovered magnitudes, which is the case for our photometric scheme 
(see section \ref{detection.photometry}). 
 It is worth noticing that at $K_{s,AB} > 25$ we actually recovered
 slightly more counts in the {\em observed} magnitude bins than we put
 in. This is caused by the fact that, in the case of a rising count slope,
 there are more faint galaxies boosted by positive noise peaks than
 bright galaxies lost on negative noise peaks. This effect is strong
 at low signal-to-noise fluxes and results in a slight excess of
 recovered counts. This is the main reason that we required little
 correction up to the detection threshold, except for the faintest
 $0.5$ mag bin centered on $K_s=26.0$, which contained false positive
 detections due to noise. After removing stars (see section
 \ref{photz.stars}), we plot in Figure \ref{fig.counts_k} 
 the raw and corrected source counts against total magnitude.

\par

\begin{figure*}[t]
\begin{center}
\includegraphics[width=0.6\textwidth]{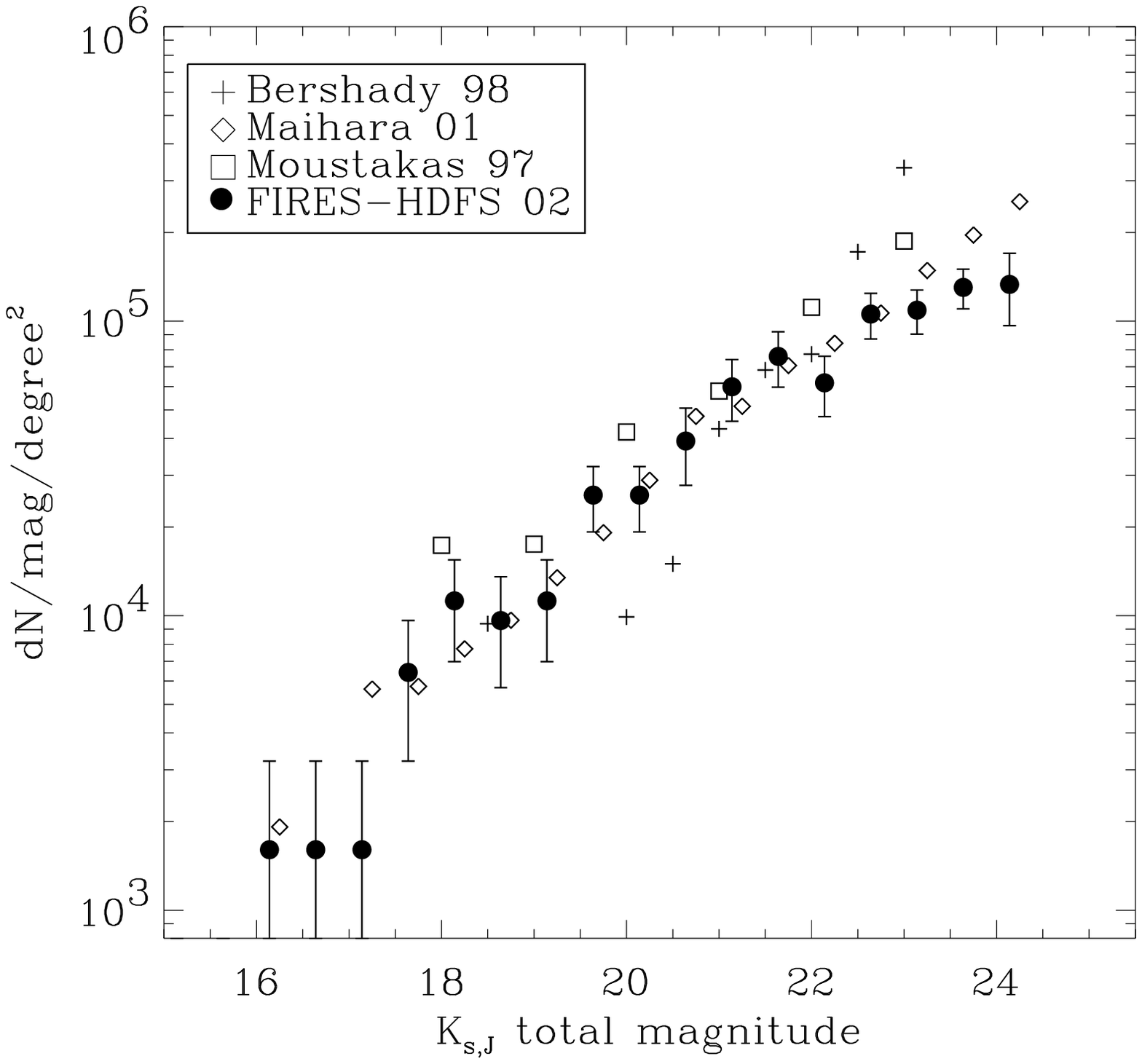}
\end{center}
\figcaption[Labbe.fig10.ps]
{FIRES $K_s-$band
galaxy counts (on the conventional Johnson system) compared to
published counts in deep $K$-band fields. The corrected counts ({\it
filled circles}) are shown for FIRES data. The \citet{Ma01} counts
have been plotted to their S/N$\sim 3$ limit. The slope at magnitudes
$K_{J} > 21$ is flatter than reported in other surveys although
straightforward comparisons are difficult, due to model-dependent
correction factors of $\sim2-3$ applied to the faintest data points
in these surveys. The nature of the scatter in count slopes is unclear
but field-to-field variations as well as different photometry and
corrections procedures likely play a role. The FIRES counts need
little correction for completeness effects or false positive
detections, except for the $K_{s,J}=24.25$
bin.\label{fig.counts_k_compare}}
\end{figure*}

Figure \ref{fig.counts_k_compare} presents a compilation of other deep
$K$-band number counts from a number of published studies. The FIRES
counts follow a d$log$(N)/dm relation with a logarithmic slope
$\alpha\approx0.25$ at $20  \lesssim K_{s,J} \lesssim 22$ (Johnson
magnitudes) and decline at fainter magnitudes to $\alpha\approx0.15$
at $22.0 \lesssim K_{s,J} \lesssim 24$. This flattening of the slope
has not been seen in other deep NIR surveys, where we emphasize that
the FIRES HDFS field is the largest and the deepest amongst these
surveys, and that only the counts in the last FIRES bin at $K_{s,J} =
24.25$ were substantially corrected. It is remarkable that the SUBARU
Deep Field count slope $\alpha\approx0.23$ of \citet{Ma01} looks
smooth compared to the HDF-S although their survey area and the raw count
statistics are slightly smaller.

Other authors \citep{Dj95,Mou97,Be98} find logarithmic counts slopes
in $K$ ranging from $0.23$ to $0.36$ over $20 \lesssim K_J \lesssim
23-24$, however the counts in the faintest bins in these surveys were
boosted by factors of $\sim2-3$, based on completeness
simulations. The origin of the faint-end discrepancies of the $K$
counts is unclear. Cosmic variance can play a role, because the survey
areas never exceed a few armin$^2$, but also differences in the used
filters ($K_s, K\prime, K$) and differences in the techniques and
assumptions used to estimate the total magnitude (see
\ref{fig.apercor}) or to correct the counts for incompleteness may be
important. Further analysis is needed to ascertain whether size-dependent
biases in the completeness correction play a role in the faint-end count
slope.

\begin{figure*}
\begin{center}
\includegraphics[width=0.8\textwidth]{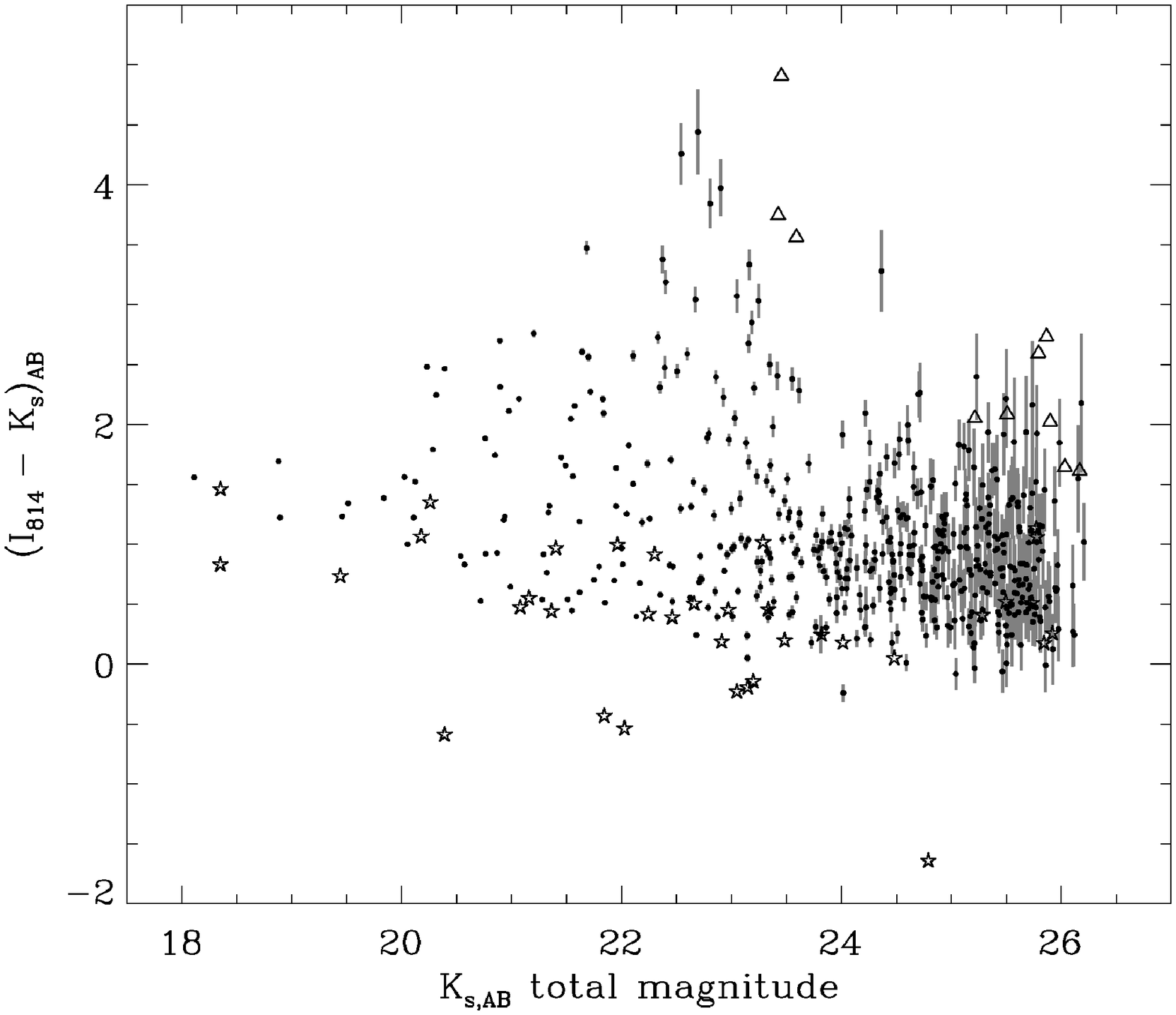}
\end{center}
\figcaption[Labbe.fig11.ps]
{\cik\ versus \vltk\ color-magnitude
relation (on the AB system) for $K_s$-selected objects in the
\hdfs. Only sources with a minimum of 20\% of the total exposure time in
all bands are included and identified stars are marked by a star
symbol. Colors are plotted with $\pm1\sigma$ error bars, and \wfi\
measurements with S/N $< 2$ ({\it triangles}) are plotted at their
$2\sigma$ confidence interval, indicating lower limits for the
colors. There are more red sources with $I_{814} - K_s > 2.6 $ at
$K\sim 23$ than at at $K\sim24$ where the $I_{814}$ is still
sufficiently deep to select them. The transformation of the $I_{814} -
K_s$ color from the AB system to the Johnson magnitude system is
$(I_{814} - K_s)_{J} = (I_{814} - K_s)_{AB} + 1.43$.
\label{fig.cm_IK-K}}
\end{figure*}

\begin{figure*}
\begin{center}
\includegraphics[width=0.8\textwidth]{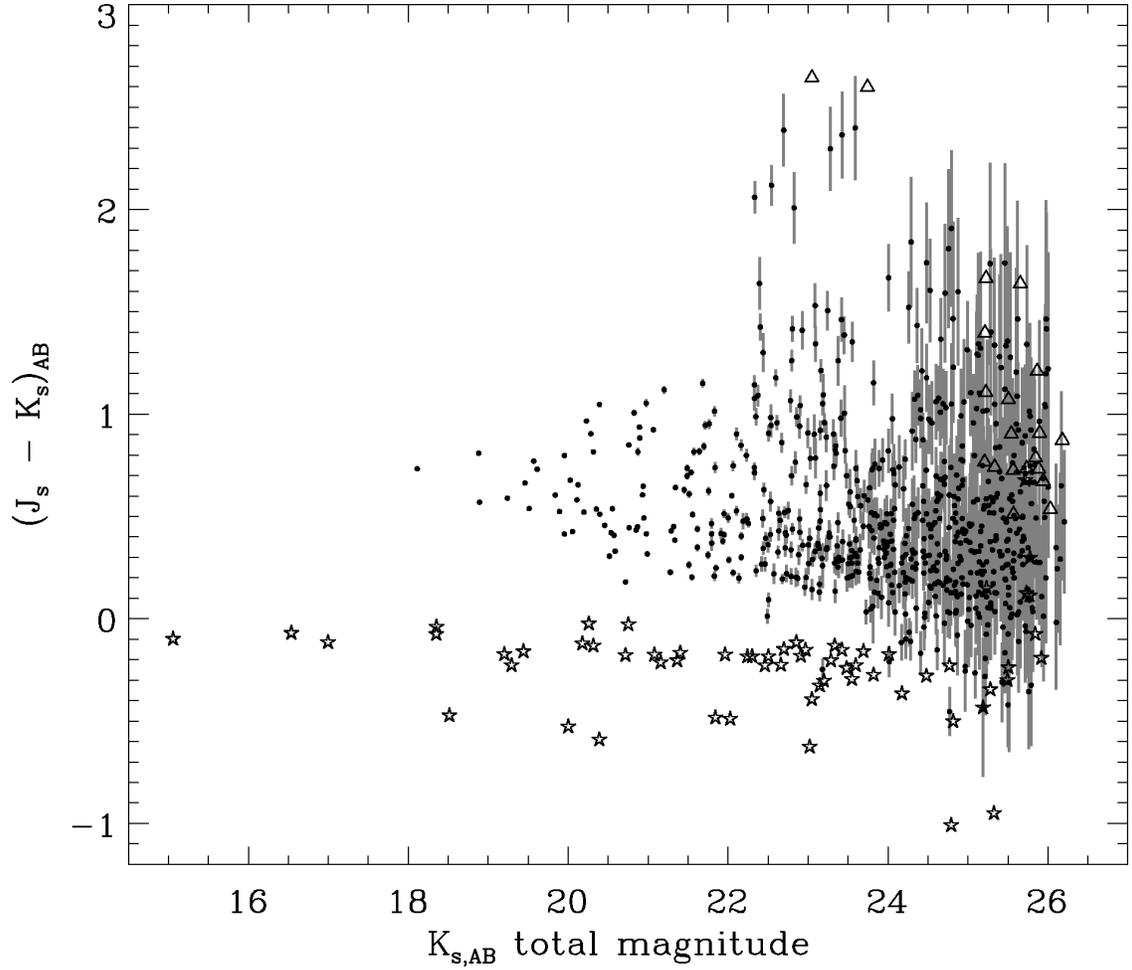}
\end{center}
\figcaption[Labbe.fig12.ps]
{Same as \myfig\ 11 for the
  $J_s-K_s$ color. Striking is the number the galaxies with very red
  NIR colors $J_s-K_s\gtrsim 1.34$ (on the AB system) or
  $J_s-K_s\gtrsim 2.3$ (Johnson). These systems have photometric
  redshifts $z > 2$ and are extremely faint in the observer's optical;
  as such they would not be selected with the U-dropout
  technique. Identified stars are well separated from redshifted
  galaxies and almost all have $J_s-K_s \leq 0$ colors. The
  transformation of the $J_s - K_s$ color from the AB system to the
  Johnson magnitude system is $(J_s - K_s)_{J} = (J_s - K_s)_{AB} +
  0.96$.  \label{fig.cm_JK-K}}
\end{figure*}

\begin{figure*}
\begin{center}
\includegraphics[width=0.8\textwidth]{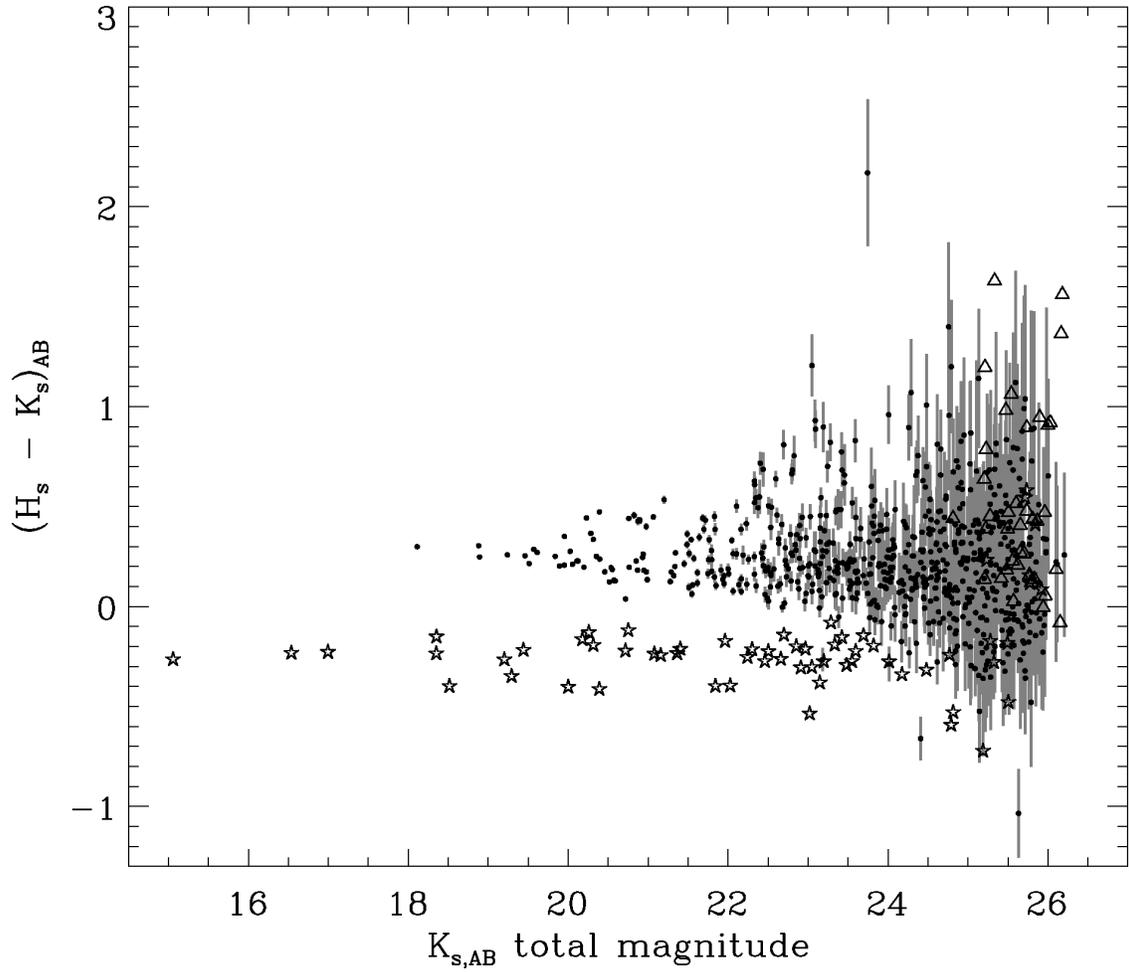}
\end{center}
\figcaption[Labbe.fig13.ps]
{Same as \myfig\ 11 for the $H-K_s$
color. One of the galaxies is extremely red with $H-K_s\approx2.2$ and
is barely visible in $J_s$ and $H$. The transformation of the $H -
K_s$ color from the AB system to the Johnson magnitude system is $(H -
K_s)_{J} = (H - K_s)_{AB} + 0.48$. \label{cm_HK-K}}
\end{figure*}

\subsection{Color-Magnitude Distributions}
\label{results.colors}

\begin{figure*}
\begin{center}
\includegraphics[width=0.8\textwidth]{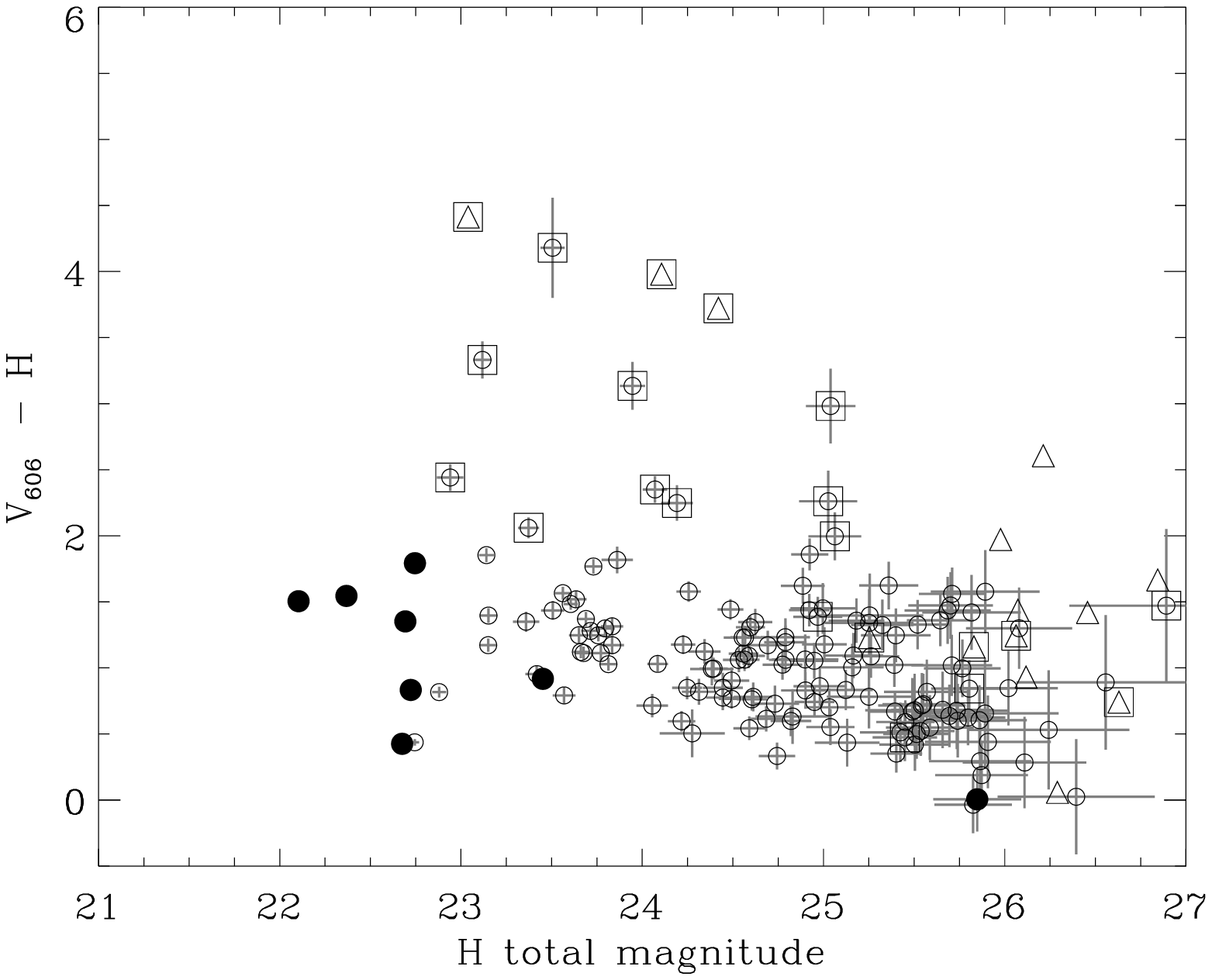}
\end{center}
\figcaption[Labbe.fig14.ps]
{$V_{606} - H$ versus $H$
color-magnitude diagram (on the AB system) for galaxies in the HDF-S
$K_s$-selected catalog with $1.95 < z_{phot} < 3.5$. Filled symbols
indicate galaxies with spectroscopy. The number of candidates for red,
evolved galaxies is much higher than in the HDF-N for a similar survey
area, as shown in a identical plot in Fig. 1 of \cite{Pa01}: we find 7
galaxies redder than $V_{606,AB} - H_{AB} \gtrsim 3$ and brighter than
$H_{AB} \lesssim 25.5$, compared to only one in the HDF-N. Galaxies with S/N $<
2$ for the $V_{606}$ measurement ({\it triangles}) are plotted at the
$2\sigma$ confidence limit in $V_{606}$, indicating a lower limit on
the $V_{606} - H$ color. The subsample of galaxies having red $(J_s -
K_s)_J > 2.3$ colors ({\it open squares}) is also shown. The
transformation of the $V_{606} - H_s$ color from the AB system to the
Johnson magnitude system is $(V_{606} - H)_{J} = (V_{606} - H)_{AB} +
1.26$. \label{fig.cm_VH-H}}
\end{figure*}

Figures \ref{fig.cm_IK-K} -- \ref{fig.cm_VH-H} show color-magnitude
diagrams of \vltk-selected galaxies in the HDF-S. The $I_{814}-K_s$
versus $K_s$ color-magnitude diagram in Figure \ref{fig.cm_IK-K} shows
a large number of extremely red objects (EROs) with $I_{814} - K_s
\gtrsim 2.6$ (on the AB system) or $(I_{814} - K_s)_J \gtrsim 4$
(Johnson). There appears to be an excess of EROs at total magnitudes
$K_{s,AB} \sim 23$ compared to magnitudes $K_{s,AB} \sim 24$. This is 
not caused by the insufficient signal-to-noise ratio in the $I_{814}$ 
measurements. In a similar diagram
for the $J_s-K_s$ color shown in Figure \ref{fig.cm_JK-K}, there is a
striking presence at the same $K_s$ magnitudes of sources with very
red $(J_s-K_s)_{AB} \gtrsim 1.34$ or $(J_s - K_s)_J \gtrsim 2.3$
colors. Such sources were also found by \citet{Sa01}, using 
shallow NIR data,  who suggested
they might be dusty starbursts or ellipticals at $z >
2$. Interestingly, any evolved galaxy with a prominent Balmer/4000\AA\
discontinuity in their spectrum, like most present-day Hubble Type
galaxies, would have such very red observed NIR colors if placed at
redshifts $z>2$. While the $(J_s - K_s)_J \gtrsim 2.3$ sources we find are
generally morphologically compact, with exceptions, we do not expect
the sources with good photometry to be faint cool L-dwarf stars
because known colors of such stars are $(J_s - K_s)_J \lesssim 2.1$
(see section \ref{photz.stars}). The photometric redshifts of all red
NIR galaxies are $z_{phot} \gtrsim 2$, but they would be missed by
ultraviolet-optical color selection techniques such as the U-dropout
method, because most of them are barely detectable even in the deepest
optical images. One bright NIR galaxy is completely undetected in 
the original WFPC2 images. The $(J_s - K_s)_J \gtrsim 2.3$ sources are studied in more
detail by \citet{Fr02} and the relative contributions of these
galaxies and U-dropouts to the rest-frame optical luminosity density
will be presented in \citet{Ru02b}.
\par

If we select sources with $1.95 < z_{phot} < 3.5$, we find clear
differences in the $V_{606}-H$ versus \vlth\ color-magnitude diagram
between our NIR-selected galaxies in the HDF-S and those of the HDF-N
(compare Figure \ref{fig.cm_VH-H} to Figure 1 of \citealt{Pa01}). Over
a similar survey area and to similar limiting depths, we find 7 
galaxies redder than $(V_{606} - H)_{AB} \gtrsim 3$ and brighter than 
total magnitude $H_{AB} \lesssim 25.5$,
compared to only one in the HDF-N.

\begin{figure*} [t]
\begin{center}
\includegraphics[width=0.6\textwidth]{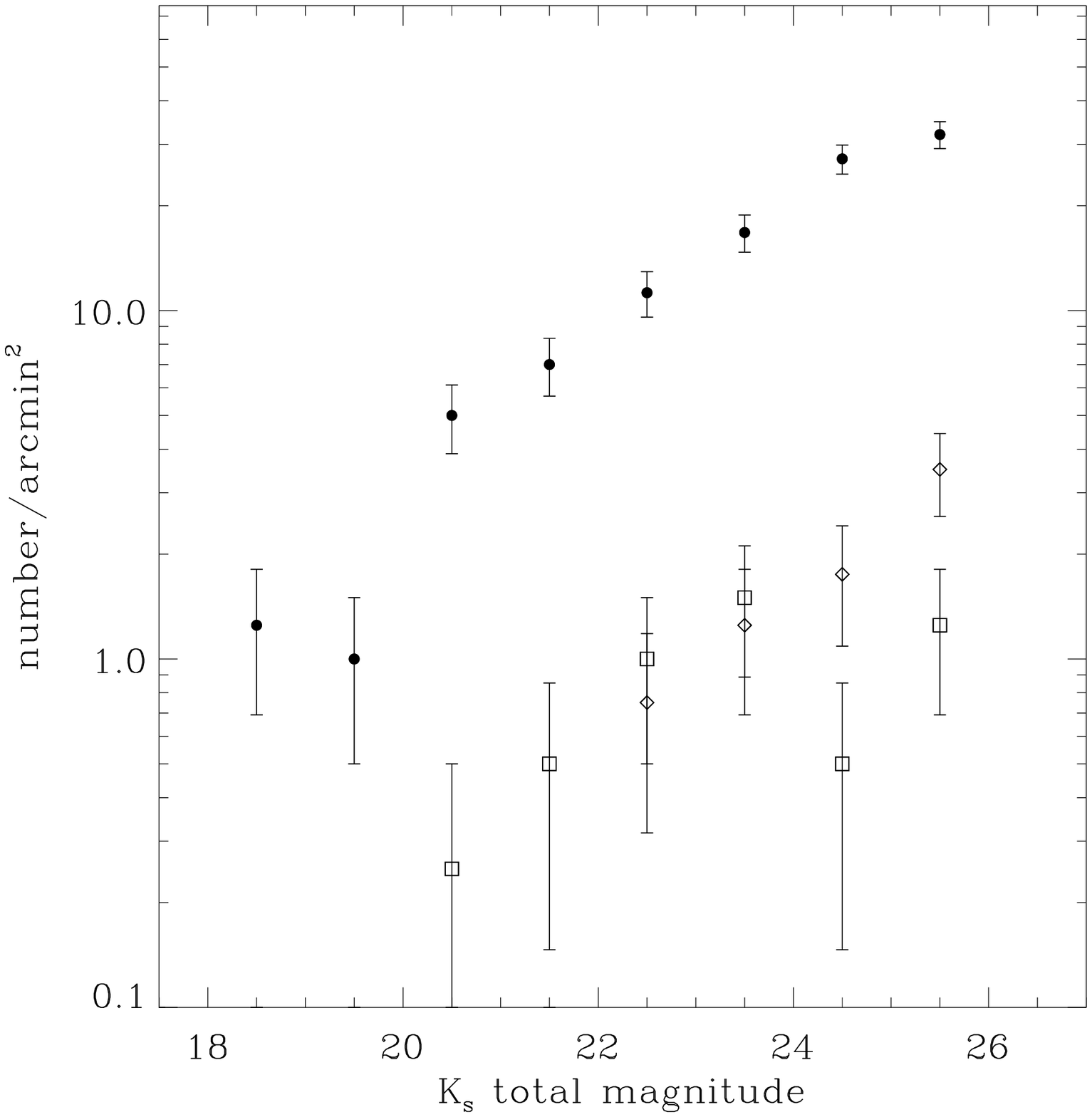}
\end{center}
\figcaption[Labbe.fig15.ps]
{The surface densities of
galaxies selected by color in the 5$\sigma$ catalog of sources of
HDF-S. Presented are galaxies with $(J_s - K_s)_J > 2.3$ (in Johnson
magnitudes) ({\it diamonds}), extremely red objects with $(I_{814} -
K_s)_{J} > 4$ ({\it squares}), and all $K_s$-selected galaxies
({\it filled points}) as a function total $K_s$-band AB
magnitude. Only sources with a minimum exposure time of 40\%, 40\% and
90\% of the total in $I_{814}, J_s$ and $K_s$ are plotted, so that the
selection in $K_s$ is uniform over the area, and the $I_{814}$ and
$J_s$ observations are sufficiently deep to prevent a bias against
objects with very red $I_{814} - K_s$ and $J_s - K_s$ colors. No
corrections to the counts have been applied. The errorbars are
poissonian and might underestimate the true uncertainty which would
also contain contributions from large scale
structure. \label{fig.counts_red}}
\end{figure*}

While the surface density of such
galaxies is not well known, it is clear that the HDF-N contains far
fewer of them than the HDF-S. It remains to be seen if this is just
field-to-field variation, or that one of the two fields is
atypical. The results of the second much larger FIRES field centered
on MS1054-03 \citep{Fo02} should provide more insight into this
issue. We note that all 7 $(V_{606} - H)_{AB} \gtrsim 3$ galaxies in the
HDF-S are also amongst the brightest 16 $(J_s - K_s)_J > 2.3$
sources. Figure \ref{fig.counts_red} shows the surface densities of
EROs and galaxies with $(J_s - K_s)_J > 2.3$ colors as function of
$K_s$-band total magnitude. The surface density of EROs peaks around
$K_{s,AB}\approx 23$ and then drops or flattens at fainter magnitudes,
contrary to the number of $(J_s - K_s)_J > 2.3$ galaxies which keeps
rising to the faintest $K_s$ total magnitudes in our catalog.

\section{Summary and Conclusions}

We have presented the results of the FIRES deep NIR imaging of the
WFPC2-field of the HDF-S obtained with ISAAC at the VLT: the deepest
ground-based NIR data available, and the deepest \vltk-band of any
field. We constructed a \vltk-selected multicolor catalog of galaxies,
consisting of 833 objects with $K_{s,AB} \lesssim 26$ and photometry
in seven-bands from $0.3$ to $2.2\mu$ for \wmincounts\ of them. These
data are available electronically together with photometric redshifts
for \nphotz\ galaxies. Our unique combination of deep optical
space-based data from the HST together with deep ground-based NIR data
from the VLT allows us to sample light redder than the rest-frame
V-band in galaxies with $z\lesssim3$ and to select galaxies from their
rest-frame optical properties, obtaining a more complete census of the
stellar mass in the high-redshift universe.  
We summarize our main findings below:
\par

\begin{itemize}

\item The $K_s$-band galaxy counts in HDF-S turn over at the faintest
magnitudes and flatten from $\alpha \approx 0.25$ at total AB
magnitudes $22 \lesssim K_s \lesssim 24$ to $\alpha \approx 0.15$ at
$24 \lesssim K_s \lesssim 26$; this is flatter than counts in
previously published deep NIR surveys, where the FIRES HDF-S field is
largest and deepest amongst these surveys. The nature of the scatter
in the faint-end counts is yet unclear but field-to-field varations as
well as different analysis techniques likely play a role.

\item The HDF-S contains 7 sources redder than $(V_{606} - H)_{AB} \gtrsim
3$ and brighter than total magnitude $H_{AB} \lesssim 25.5$ at
photometric redshifts $1.95 < z_{phot} < 3.5$, while such galaxies
were virtually absent in the HDF-N. They are much redder than regular
U-dropout galaxies in the same field and are candidates for relatively
massive, evolved systems at high redshift. The difference with the
HDF-N might just reflect field-to-field variance, calling for more
observations to similar limits with full optical-to-infrared
coverage. Results from the second and larger FIRES field centered on
MS1054-03 \citep{Fo02} should provide more insight into this
issue. 

\item We find substantial numbers of red galaxies with $(J_s - K_s)_J
> 2.3$ that have photometric redshifts $z_{phot} > 2$. These galaxies
would be missed by ultraviolet-optical color selection techniques such as
the U-dropout method because most of them are barely detectable even
in the deepest optical images. The surface densities of these sources
in our field keeps rising down to the detection limit in $K_s$, in
contrast to the number counts of EROs which peak at $K_{s,AB}\sim 23$
and then drop or flatten at fainter magnitudes.

\end{itemize}
\par

The results of the HDF-S presented in this paper demonstrate the
necessity of extending optical observations to near-IR wavelengths for
a more complete census of the early universe. Our deep $K_s$-band data
prove invaluable for they probe well into the rest-frame optical at $2
< z < 4$, where long-lived stars may dominate the light of
galaxies. We are pursuing follow-up programs to obtain more
spectroscopic redshifts needed to confirm the above results. 
Updates on the FIRES programme and access to the reduced images and 
catalogues can be found at our website \fireswww.

\acknowledgements

The data here presented have been obtained as part of an ESO Service
Mode program. We would very much like to thank the ESO staff for their
kind assistance and enormous efforts in taking these data and making them
available to us. This project would not be possible without their
dedicated work. This work was supported by a grant from the 
Netherlands Organization for Scientific Research. We would like to thank
the Lorentz Center of Leiden University for its hospitality during 
several workshops. The comments of the referee helped to improve the
paper.

\clearpage

\clearpage
\begin{deluxetable}{ccrcc}
\tablecaption{Summary of the HDF-S Observations} 
\tablewidth{0pt} 
\tablehead{\colhead{Camera} & \colhead{Filter}& \colhead{Number of} & \colhead{Integration Time}  & \multicolumn{1}{c}{Image Quality\tablenotemark{1}} \\
\colhead{} & \colhead{} & \colhead{Exposures} & \colhead{(h)} & \colhead{(arcsec)}}
\startdata
WFPC2 & $F300W$   &  102  &  36.8  &  0\farcs16 \\
WFPC2 & $F450W$  &  51   &   28.3 &  0\farcs14 \\
WFPC2 & $F606W$  &  49   &   27.0  &   0\farcs13 \\
WFPC2 & $F814W$  &  56   &   31.2 &   0\farcs14 \\
ISAAC & $J_s$   & 1007  & 33.6 &  0\farcs45  \\
ISAAC & $H$     & 968   & 32.3  &  0\farcs48 \\
ISAAC & $K_s$   & 2136  & 35.6  &  0\farcs46 \\
\enddata
\tablenotetext{1}{The full width at half-maximum of the best-fitting Gaussian.}
\label{table.HDFS.summary} 
\end{deluxetable}

\begin{deluxetable}{ccc}
\tablecaption{Zero Points for the HDF-S} 
\tablewidth{0pt} 
\tablehead{\colhead{} & \multicolumn{2}{c}{Zero Point} \\
\colhead{Data Set} & \colhead{Johnson mag} & \colhead{AB mag} }
\startdata
$U_{300}$  &  19.43  &  20.77  \\
$B_{450}$  &  22.01  &  21.93 \\ 
$V_{606}$  &  22.90  &  23.02   \\
$I_{814}$  &  21.66  &  22.09   \\
$J_s$   &  24.70  &  25.60   \\
$H$     &  24.60  & 25.98   \\
$K_s$   &  24.12  & 25.98    \\
\enddata
\label{table.ZP.summary} 
\end{deluxetable}

\begin{deluxetable}{ccrcc}
\tablecaption{Background noise in the HDF-S images} 
\tablewidth{15cm} 
\tablehead{\colhead{Data Set\tablenotemark{1}} & \colhead{rms background variation\tablenotemark{2}} & \colhead{a\tablenotemark{3}} & \colhead{b\tablenotemark{3}} & \colhead{1$\sigma$ sky noise limit\tablenotemark{4}}}
\startdata
$U_{300}$  &  1.34e-05 & 2.51  & 0.38  & 29.5\\
$B_{450}$  &  1.88e-05 & 2.65  & 0.40  & 30.3\\
$V_{606}$  &  3.80e-05 & 2.49  & 0.39  & 30.6\\
$I_{814}$  &  2.79e-05 & 2.59 & 0.39  & 30.0\\
$J_s$   &   0.0069  &  1.46 &   0.047 &  28.6\\
$H$     &   0.0165   &  1.43  &   0.044 &  28.1\\
$K_s$   &   0.0163   &   1.49 &  0.038 & 28.1\\
\enddata
\tablenotetext{1}{The images are all at the $0\farcs119$~pixel$^{-1}$ scale. The WFPC2 are 3x3 block summed, and all images are smoothed to match the image quality of the $H$-band.}
\tablenotetext{2}{Pixel-to-pixel rms variations (in instrumental counts per second) as measured directly in empty parts of the registered convolved images which were used for photometry.}
\tablenotetext{3}{Best-fit parameters of Eq. \ref{error.scaling} which gives the effective rms variation of the background as a function of linear size of the aperture.}
\tablenotetext{4}{The $1\sigma$ sky noise limit in a $0\farcs7$ circular diameter aperture ($\approx 0.4$ arcsec$^2$) in AB magnitudes using Eq. \ref{error.scaling}.}
\label{table.background}
\end{deluxetable}

\end{document}